\PassOptionsToPackage{table}{xcolor}
\documentclass[acmsmall,nonacm]{acmart}
\usepackage{amsmath,amsfonts}
\usepackage{algorithmic,algorithm}
\usepackage{graphicx}
\usepackage{textcomp}
\usepackage{booktabs}
\usepackage{subcaption}
\usepackage{enumitem}
\usepackage{pifont}
\usepackage{epstopdf}
\usepackage[utf8]{inputenc}
\usepackage{tablefootnote}
\usepackage{ifthen}

\usepackage{url}
\usepackage{multirow}

\usepackage{dblfloatfix}

\def\BibTeX{{\rm B\kern-.05em{\sc i\kern-.025em b}\kern-.08em
    T\kern-.1667em\lower.7ex\hbox{E}\kern-.125emX}}

\setcopyright{rightsretained}
\acmJournal{TECS}
\acmYear{2025}
\acmVolume{1}
\acmNumber{1}
\acmArticle{1}
\acmMonth{9}
\acmDOI{}

\newcommand{\bh}[1]{\vspace{3pt}\noindent \textbf{#1} }

\newboolean{comment_bullets}
\setboolean{comment_bullets}{False}

\begin{document}

\title{THERMOS: Thermally-Aware Multi-Objective Scheduling of AI Workloads on Heterogeneous Multi-Chiplet PIM  Architectures}

\author{Alish Kanani}
\email{ahkanani@wisc.edu}
\author{Lukas Pfromm}
\email{pfromm@wisc.edu}
\affiliation{%
  \institution{University of Wisconsin–Madison}
  \city{Madison}
  \state{Wisconsin}
  \country{USA}
}

\author{Harsh Sharma}
\author{Janardhan Rao Doppa}
\author{Partha Pratim Pande}
\affiliation{%
  \institution{Washington State University}
  \city{Pullman}
  \state{Washington}
  \country{USA}
}

\author{Umit Y. Ogras}
\affiliation{%
  \institution{University of Wisconsin–Madison}
  \city{Madison}
  \state{Wisconsin}
  \country{USA}
}


\thanks{This work was partially supported by the Intel CAD SRS program.}



\begin{abstract}

Chiplet-based integration enables large-scale systems that combine diverse technologies, enabling higher yield, lower costs, and scalability, making them well-suited to AI workloads.
Processing-in-Memory (PIM) has emerged as a promising solution for AI inference, leveraging technologies such as ReRAM, SRAM, and FeFET, each offering unique advantages and trade-offs.
A heterogeneous chiplet-based PIM architecture can harness the complementary strengths of these technologies to enable higher performance and energy efficiency.
However, scheduling AI workloads across such a heterogeneous system is challenging due to competing performance objectives, dynamic workload characteristics, and power and thermal constraints. 
To address this need, we propose THERMOS, a thermally-aware, multi-objective scheduling framework for AI workloads on heterogeneous multi-chiplet PIM architectures. 
THERMOS trains a \textit{single multi-objective reinforcement learning (MORL) policy} that is capable of achieving Pareto-optimal execution time, energy, or a balanced objective at runtime, depending on the target preferences. 
Comprehensive evaluations show that THERMOS achieves up to 89\% faster average execution time and 57\% lower average energy consumption than baseline AI workload scheduling algorithms with only 0.14\% runtime and 0.022\% energy overhead. 
\end{abstract}



\maketitle

\vspace{-4pt}
\section{Introduction} \label{sec:introduction}
Semiconductor roadmaps emphasize an unprecedented demand for memory and processing over the next decade~\cite{src_report, IRTS2015}. 
Since a single monolithic die cannot satisfy this need, heterogeneous integration through advanced packaging has emerged as a promising solution to interconnect tens to hundreds of chiplets with significant cost and performance benefits~\cite{packaging_roadmap}. 
Chiplet-based systems enable cost-effective scaling since yields of individual (smaller) chiplets are significantly higher than those of large monolithic designs~\cite{graening2024cost, stow2016costAnalysis}.
This approach enables designers to integrate multiple chiplets fabricated with different technologies on an interposer.

Conventional von Neumann architectures face memory bottlenecks due to rapidly increasing computational requirements of data-intensive AI/ML applications and the accompanying data movement.
PIM has emerged as a new paradigm to minimize data movement by performing computations in memory where the model weights are stored~\cite{jia2020programmable,sinagil2021cim,wan202033cim,wang2023charge,kim2021colonnade}. 
Crossbar arrays, widely used for PIM, enable highly efficient matrix-vector multiplication (MVM), which forms the core of deep learning (DL) workloads. 
In analog-domain MVM, voltages representing the input values are applied to the rows, and the resulting currents flowing through each column are accumulated to produce the outputs.
These accumulated analog currents are then converted back into digital values using analog-to-digital converters (ADCs)~\cite{neurosim}.
Prominent PIM architectures use SRAM cells or non-volatile memory (NVM) devices such as Resistive Random Access Memory (ReRAM), Spin-Transfer Torque Magnetic RAM (STT-MRAM), and Ferroelectric Field-Effect Transistor memory (FeFETs)~\cite{shafiee2016isaac,roy2020memory,soliman2023first,gajaria2024chime,krestinskaya2024neural}.

While NVM-based PIM systems enable fast and energy-efficient MVM operations, these systems suffer from increased sensitivity to high temperatures. 
For example, ReRAM-based implementations can efficiently perform dense computations due to its compact 1T-1R memory cell structure. 
However, ReRAM is more sensitive to high temperatures, which can impact reliability~\cite{liu2019hr,meng2021temperature}. 
In contrast, SRAM-based PIM uses larger eight-transistor (8T) memory cells, which leads to higher area overhead but offers better stability~\cite{jaiswal20198t}.

\begin{figure}[t]
\centering
    \begin{subfigure}[b]{0.69\textwidth}
        \includegraphics[width=\linewidth]{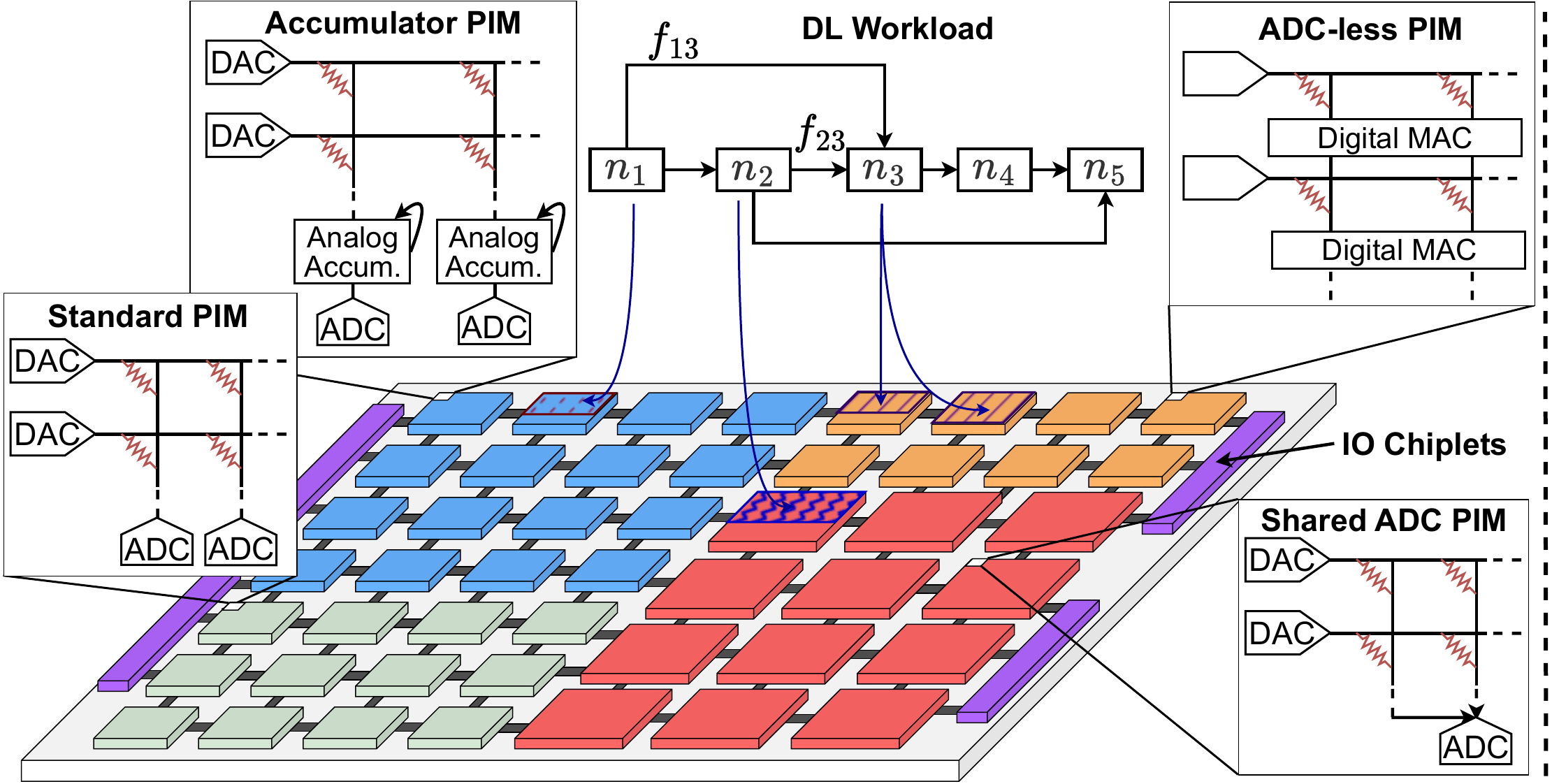}
        \caption{}
        \vspace{-10pt}
        \label{fig:pim_system}
    \end{subfigure}
    \hfill
    \begin{subfigure}[b]{0.27\textwidth}
        \includegraphics[width=\linewidth]{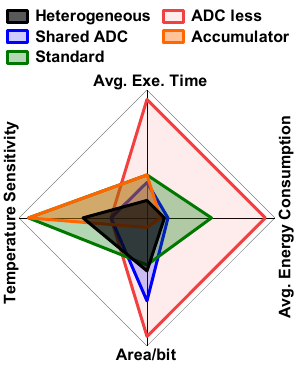}
        \caption{}
        \label{fig:radar_plot}
    \end{subfigure}
    \caption{Overview of the heterogeneous multi-chiplet PIM architecture. (a) A heterogeneous PIM system with four chiplet clusters (standard, shared ADC, ADC-less, accumulator). The neural layers of the DL workload are first assigned to a cluster and then to a chiplet within that cluster. (b) A radar plot comparing execution time, energy, memory density, and thermal sensitivity of homogeneous vs. heterogeneous PIM architectures, based on simulated systems with equal processing area. }
    \vspace{-10pt}
    \label{fig:overview}
\end{figure}

Integrating different PIM implementations within a single heterogeneous system can significantly improve the overall performance by leveraging the strengths of each PIM type~\cite{krishnan2022biglittle}.
Since each technology offers unique advantages, their heterogeneous integration in a multi-chiplet system provides significant benefits over homogeneous systems, as illustrated in Figure~\ref{fig:overview}.
This work considers DL inference applications (AI workloads) streaming to a heterogeneous chiplet system comprising of four promising PIM types:
(1) Standard ReRAM-based chiplets~\cite{neurosim, krishnan2021siam, shafiee2016isaac}, 
(2) SRAM-based chiplets with shared ADCs~\cite{jia2020programmable}, 
(3) ADC-less SRAM-based chiplets~\cite{kim2021colonnade,saxena2022towards}, and 
(4) ReRAM-based chiplets that employ analog accumulators to add input across multiple cycles~\cite{wan202033cim}. 

The goal of designing these large-scale heterogeneous systems is to co-optimize throughput (the number of DL inference tasks processed per unit of time) and energy efficiency. 
However, the dense integration of chiplets within a single package introduces significant challenges for power dissipation, thermal management, and reliability~\cite{kim2023thermomechanical,mounce2016chiplet,park2024thermal}.
Overcoming these challenges while achieving optimal performance and energy efficiency requires optimizing multiple aspects of the system, including application-to-chiplet scheduling and inter-chiplet communication latency.
Hence, there is a critical need for multi-objective scheduling methodologies for large-scale PIM architectures with hundreds of chiplets.
In this paper, we focus on runtime DL inference workload scheduling to co-optimize execution time and system-wide energy consumption under thermal constraints by maximizing the throughput of the heterogeneous multi-chiplet PIM system.


This paper proposes THERMOS, a \textit{multi-objective, thermally-aware scheduling framework} for running DL workloads on heterogeneous multi-chiplet PIM architectures. 
We formulate the DL workload scheduling task as a constrained multi-objective optimization problem that aims to jointly minimize execution time and maximize energy efficiency while satisfying thermal limits. 
THERMOS solves this problem using a hierarchical scheduling approach involving two scheduling levels. At the first level, a multi-objective reinforcement learning (MORL) policy selects the appropriate type of PIM chiplet (i.e., a PIM chiplet cluster) for each neural layer in the DL workload. 
Our MORL-based approach trains \textit{a single policy that dynamically adapts to a given preference vector at runtime}, achieving Pareto-optimal performance in terms of execution time and energy consumption. We consider three key preferences: minimizing execution time, minimizing energy consumption, and balancing both objectives.
We emphasize that \textit{THERMOS delivers a single policy} that takes these preferences as input at runtime and delivers Pareto-optimal objectives, considering the thermal impacts of its scheduling actions.
At the second level, a proximity-driven algorithm assigns neural layers to specific chiplets within the cluster selected by the MORL technique, aiming to minimize communication costs. 

We evaluate THERMOS across four different network-on-interposer (NoI) architectures and compare it against three state-of-the-art scheduling algorithms while executing DL inference workloads.
THERMOS achieves up to 36\% faster execution, 22\% lower energy, and 36\% lower EDP than the scheduling policy used in Simba~\cite{simba}, a popular chiplet architecture. 
It delivers up to 89\% faster execution, 57\% lower energy, and 113\% lower EDP compared to the scheduling strategy of the recent heterogeneous Big-Little~\cite{krishnan2022biglittle} architecture. 
Finally, it achieves up to 58\% faster execution, 26\% lower energy, and 104\% lower EDP than RELMAS~\cite{blanco2024deep}, a state-of-the-art RL-based scheduling approach.
We also evaluate its overhead on an Nvidia Jetson Xavier NX platform~\cite{nvidia2020jetson}, showing a negligible 0.14\% runtime and 0.022\% energy overhead per workload.


The major contributions of this work are as follows:
\begin{itemize}[leftmargin=*]
    \item This work is the \textit{first to demonstrate thermally-aware multi-objective workload scheduling} for heterogeneous multi-chiplet PIM architectures.
    \item We present a single MORL policy that can receive a preference vector at runtime and optimize for performance, energy consumption, or a balanced objective.
    \item Extensive evaluation on various DL inference workloads and multiple heterogeneous chiplet systems demonstrates that THERMOS achieves the highest throughput for any given system and outperforms the state-of-the-art techniques~\cite{krishnan2022biglittle,blanco2024deep,simba} with up to 89\% lower execution time and 55\% less energy consumption.
\end{itemize}

The rest of this paper is organized as follows: Section~\ref{sec:related_work} reviews related work, Section~\ref{sec:overview} outlines DL workload characteristics and THERMOS. Section~\ref{sec:approach} formulates the scheduling problem and presents our scheduler. Section~\ref{sec:experimental_results} presents a comprehensive evaluation. Finally, Section~\ref{sec:conclusion} summarizes our contributions. 
 
\section{Related Work} \label{sec:related_work}

Data movement is a primary source of energy consumption and latency in computing systems. PIM techniques address this problem by maintaining deep neural network weights on chip to enable faster and more energy-efficient execution of DL inference workloads. 
ISAAC~\cite{shafiee2016isaac} shows how PIM crossbars can be assembled into chips capable of dramatically increasing the throughput and energy efficiency of DL inference. 
Previous research has also investigated different PIM architectures to extract additional performance and to optimize data reuse~\cite{jia2020programmable, sinagil2021cim, wan202033cim, wang2023charge, kim2021colonnade,krestinskaya2024neural}. 
Although PIM is naturally suited for improving performance in memory-bound workloads, several studies show that it also benefits DL inference, especially when memory access dominates latency and energy~\cite{chen2016eyeriss,jouppi2017datacenter,jacob2018quantization,kim2016neurocube,seshadri2017ambit}. 
This is particularly evident in intermediate feature map processing, even when on-chip buffers are present~\cite{chen2016eyeriss,jouppi2017datacenter}. 
Quantized DNNs (e.g., INT8/INT4) further increase the memory-to-compute ratio per MAC operation, making them highly amenable to PIM~\cite{jacob2018quantization}. 
Recent designs like Neurocube~\cite{kim2016neurocube} and Ambit~\cite{seshadri2017ambit} report competitive throughput (0.9–1.15 TOPS/$mm^2$) and superior energy efficiency (as low as 1.1 pJ/MAC) compared to commercial NPUs (~4 TOPS/$mm^2$, 1.5–5 pJ/MAC)~\cite{lee2018unpu}. 
Moreover, while early CNN layers are compute-heavy, later layers remain memory-bound, such as fully connected and depthwise convolutions.


Chiplet-based systems have become increasingly popular due to their flexibility and low manufacturing costs~\cite{graening2024cost}.
However, they can suffer from lower performance compared to monolithic systems due to inter-chiplet data movement. 
Chiplets are connected to each other through the interposer and form a NoI. 
The NoI significantly impacts the amount of latency incurred during communication and, therefore, the system performance.
Hence, prior work has explored the design of efficient NoI, creating topologies optimized for different workloads and system types \cite{kite, krishnan2021siam, sharma2023florets, hexamesh, sharma2022swap}. 
HexaMesh creates an NoI topology with six links per chiplet to enable high bisection bandwidth, increasing the system's overall performance~\cite{hexamesh}. 
Similarly, Kite introduces a set of system topologies with long-range links designed to decrease the total hop count between chiplets~\cite{kite}. 
While these approaches are designed for general workloads, DL workloads have predictable data flow patterns that can be exploited by the NoI to improve performance and energy-efficiency. 
Floret, a data flow-aware topology, is optimized for the communication between layers in DL workloads~\cite{sharma2023florets}. 
\textit{We evaluate the proposed THERMOS framework on all of these architectures to demonstrate its applicability to any NoI topology}. 

Workload scheduling is crucial for maximizing performance in a given system~\cite{khdr2024multi,zhong2025maco,du2023accelerating,odema2024scar,senapati2023tmds}. 
SIMBA, a homogeneous chiplet-based system designed for DL workloads~\cite{simba}, employs a nearest-neighbor scheduling strategy to map the neural network layers to chiplets.
This scheduling strategy aims to minimize the total amount of communication latency incurred during DL inference by placing data producers and consumers on spatially nearby chiplets. 
NN-Baton uses the same system configuration as SIMBA and introduces an automated tool for workload to chiplet scheduling, which accounts for the multiple levels of hardware within a chiplet and generates a scheduling strategy using an analytical methodology~\cite{tan2021nnbaton}.
Similarly, MAGMA develops a DL workload to chiplet scheduling method but instead focuses on the case where several workloads arrive and must be mapped in parallel~\cite{kao2022magma}. 
These existing approaches are designed for homogeneous chiplet-based systems, but scheduling to heterogeneous systems requires new methods.
In Big-Little Chiplets, a heterogeneous system is created with big and little chiplets, where big chiplets contain more and larger PIM crossbars than small chiplets~\cite{krishnan2022biglittle}. 
To take advantage of the data flow patterns in DL workloads, earlier layers with fewer weights are mapped to Little chiplets, and later layers are mapped to Big chiplets. 
Similarly, RELMAS is an RL-based algorithm designed for real-time scheduling of DL workloads in heterogeneous multi-accelerator systems~\cite{blanco2024deep}. 
It employs a heterogeneous system consisting of Eyeriss and Simba-style tiles~\cite{chen2016eyeriss,simba}. 
Table~\ref{tab:system_comparison} provides a detailed qualitative comparison of existing scheduling methods. 
Detailed quantitative evaluation of THERMOS and comparison of its performance against Simba, Big-Little, and RELMAS are presented in Section~\ref{sec:experimental_results}.
{\em While these approaches can maximize performance, none of them are thermally-aware, leading to compromised performance if overheating occurs.} 

Thermal bottlenecks are already limiting the performance of computing systems implemented on monolithic chips~\cite{sartor2020hilite}. 
This issue is magnified in chiplet-based systems due to the close integration of multiple dies.
To identify thermal issues at the design time, prior work has created simulators capable of thermal modeling the complex geometry of chiplet-based systems \cite{hotspot,3D_ICE,pact,pfromm2024mfit}. 
Prior work has also investigated methods to reduce these bottlenecks at system design time~\cite{thermalChipletOrganizationDarkSilicon, tap_2.5d}. 
These methods modify the placement and spacing of chiplets on the interposer to reduce hotspots, thereby increasing potential performance. 
However, they are limited to design time and cannot mitigate runtime thermal bottlenecks (for example, thermal runaway). 
Hence, active approaches during runtime are required to manage the temperature of these systems. 

\begin{table}[t]
\centering
\caption{Comparison of systems created for DL inference.}\label{tab:system_comparison}
\vspace{-5pt}
\resizebox{\textwidth}{!}{%
\begin{tabular}{@{}l|c|c|c|c|c@{}}
    \toprule
    \textbf{Work} & \textbf{Heterogeneous} & \textbf{PIM} & \textbf{Scheduling Method} & \textbf{Multi-Objective}  & \textbf{Thermally-Aware}\\ \midrule
    Big-Little~\cite{krishnan2022biglittle} & Chiplet Size & N & Heuristic & N & N \\
    SIAM~\cite{krishnan2021siam} & N & Y & Heuristic & N & N\\
    SIMBA~\cite{simba} & N & N & Heuristic & N & N\\
    RELMAS~\cite{blanco2024deep} & SIMBA+Eyeriss & N & RL & N & N \\
    \textbf{THERMOS} & \textbf{PIM Architecture and Size} & \textbf{Y} & \textbf{Multi objective RL + Heuristic} & \textbf{Y} & \textbf{Y}\\
    \bottomrule
\end{tabular}%
}
\vspace{-12pt}
\end{table}

Multiple approaches exist to manage the temperature of a system during runtime. 
Dynamic voltage-frequency scaling (DVFS) has been widely applied to CMOS-based systems to adjust the amount of power consumption, and therefore heat generation, of a chip \cite{TODAES-IL,mandal2019dynamic,kim2017imitation,zanini2009multicore,bhat2017algorithmic}. 
However, voltage and frequency cannot be modified in some computing technologies, including PIM, as their analog circuits are designed for specific voltage levels to ensure optimal accuracy.
Therefore, alternative methods are needed to manage the system's temperature. Thermally-aware task scheduling can meet this critical need. 
By managing the mapping of workloads to the system, the temperature of the system can be prevented from crossing a set threshold. 
The HR3AM framework addresses these challenges by enhancing the thermal resilience of ReRAM-based neural network accelerators~\cite{liu2019hr}.
Additionally, the WRAP framework proposes a thermally-aware weight remapping strategy to manage the thermal implications of neural network processing in ReRAM-based systems~\cite{chen2022wrap}.
However, these existing approaches are not designed for the unique compute-communication structure of chiplet-based systems. 
In contrast to existing scheduling approaches, THERMOS introduces a hierarchical scheduling framework designed specifically for chiplet-based systems, which maximizes performance while preventing the system temperature from violating the specified constraints.

\section{Preliminaries and Overview}\label{sec:overview}
This section first introduces the DL inference workloads and their representation in this work. 
Next, it describes the heterogeneous multi-chiplet PIM architecture and provides an overview of the multi-objective RL approach used for dynamically scheduling the DL workloads on this heterogeneous multi-chiplet PIM system. 

\subsection{Deep Neural Network Workload Characteristics}\label{ssec:dcg_arcg}
DL workloads follow a structured, layered computation model. 
Matrix multiplication-heavy computation is localized within each neural layer, while communication occurs through activations that follow a deterministic pattern.
A neural layer refers to any computation-intensive component, such as a fully connected, convolutional, or attention layer, which adheres to a similar computation-communication model.
Therefore, once the DL model is defined, the computation requirements, communication pattern, and amount of data transfer between layers can be determined. 
Using these characteristics, we characterize DL workload as a graph:

\bh{Definition 1: DL Characterization Graph (DCG)}
is a directed graph $G_{\text{DCG}}(\mathcal{N}, \mathcal{F})$, where $\mathcal{N}$ is the set of vertices and $\mathcal{F}$ is the set of data flows between each vertex due to activations.
\begin{itemize}[leftmargin=*]
    \item Each vertex $n_i \in \mathcal{N}$ represents a neural layer and encodes its computational requirements as a tuple $(w_i, o_i)$, where $w_i$ denotes the memory required for weights, and $o_i$ represents the total multiply-accumulate (MAC) operations needed for computation. Consequently, there are $N = |\mathcal{N}|$ layers.
    \item Each arc $f_{ij} \in \mathcal{F}$ captures the total data volume from layer $i$ to layer $j$ due to the activations from layer $i$. 
\end{itemize}
After the neural layers $\mathcal{N}$ of the DL model are scheduled to the target architecture, each layer sequentially processes a stream of input frames $I$, such as images, as explained in Section~\ref{ssec:overview}. 

\subsection{Heterogeneous Multi-Chiplet PIM Architecture} \label{ssec:heterogeneous_architecture}
Various PIM architectures based on ReRAM and SRAM have been proposed for DL inference, each offering distinct advantages and limitations in terms of performance, energy, and thermal characteristics. 
Rather than relying on a single PIM technology, a heterogeneous system that integrates multiple PIM implementations can achieve superior overall performance by leveraging the strengths of each type. 
As shown in Figure~\ref{fig:radar_plot}, the heterogeneous design delivers better overall trade-offs than any homogeneous baseline.
These results highlight the practical advantage of chiplet diversity in heterogeneous architectures, enabling improved performance and energy efficiency compared to uniform PIM deployments.
This work considers four such PIM implementations described here. 


\begin{enumerate}[leftmargin=*]
\item \textbf{Standard PIM} approach utilizes ReRAM-based macros, where inputs are processed as 1-bit streaming data, and each column is connected to individual ADCs~\cite{neurosim}.

\item \textbf{Shared ADC PIM} architecture employs shared ADCs across crossbar columns to save area and energy~\cite{jia2020programmable}. 
The design requires fewer ADCs by summing analog outputs across different columns, reducing both area and energy consumption. 

\item \textbf{ADC-less PIM} architecture, also known as near-memory computing (NMC), replaces analog-domain MAC operations with fully digital processing. 
Instead of performing computations in the analog domain, each memory cell is directly connected to a digital MAC unit, eliminating the need for ADCs~\cite{kim2021colonnade}. 

\item \textbf{Accumulator-Based PIM} architecture uses analog accumulators to sum inputs across cycles before converting them into the digital domain, thereby reducing the frequency of energy-intensive ADC operations~\cite{wan202033cim}. 
\end{enumerate}

Since different PIM implementations rely on different fabrication technologies, a heterogeneous system on a monolithic chip is not feasible. 
Therefore, this work considers heterogeneous chiplet-based architectures that integrate multiple PIM chiplets. 
Each chiplet is fabricated for a specific PIM implementation and is interconnected using an NoI. 
Each chiplet also includes a buffer for storing activations and the results of MAC operations, along with peripheral circuits for implementing nonlinear functions and an on-chip network for internal connectivity~\cite{neurosim,krishnan2021siam}. 
The integrated system is a standalone platform, incorporating I/O chiplets for external connectivity to a host system and external memory, as illustrated in Figure~\ref{fig:pim_system}. 

\bh{Definition 2: Multi-Chiplet Architecture Characterization Graph (ACG)}
is a directed graph $G_{\text{ACG}}(\mathcal{C}, \mathcal{L})$, 
where $\mathcal{C}$ is the set of chiplets in the architecture and $\mathcal{L}$ represent the communication links interconnecting them.
\begin{itemize}[leftmargin=*]
\item Each vertex $ c_i \in \mathcal{C} $ corresponds to a chiplet, which is characterized by the tuple  $(v_i, M^{cap}_i, M_i(t), T_i(t), T^{max}_i)$
    \begin{itemize}
    \item $v_i$ denotes the specific PIM type, where $v_i \in \mathcal{V} = $ \{\textit{standard, shared ADC, ADC-less, accumulator}\}
    \item $M^{cap}_i$ defines the total memory available in crossbars for DL model weights
    \item $M_i(t)$ is the available memory at time $t$. It varies based on workload execution and scheduling decisions
    \item $T_i(t)$ denotes the highest temperature observed on chiplet $c_i$ at time $t$
    \item $T^{max}_i$ is the maximum allowed temperature for chiplet type $v_i$
    \end{itemize}
\item The arcs $l_{ij} \in \mathcal{L}$ represent communication links between chiplets $c_i,~c_j~\in \mathcal{C}$. Each arc  $l_{ij}$ is characterized by latency and energy per bit based on the NoI topology that interconnects the chiplets.
\end{itemize} 
Chiplets are logically clustered by PIM type $\mathcal{V}$, ensuring that all chiplets within a cluster have similar compute capabilities, total available memory in crossbars, and a common maximum allowable temperature.

\begin{figure}[b]
\centering
    \begin{subfigure}[b]{0.7\textwidth}
        \includegraphics[width=1\linewidth]{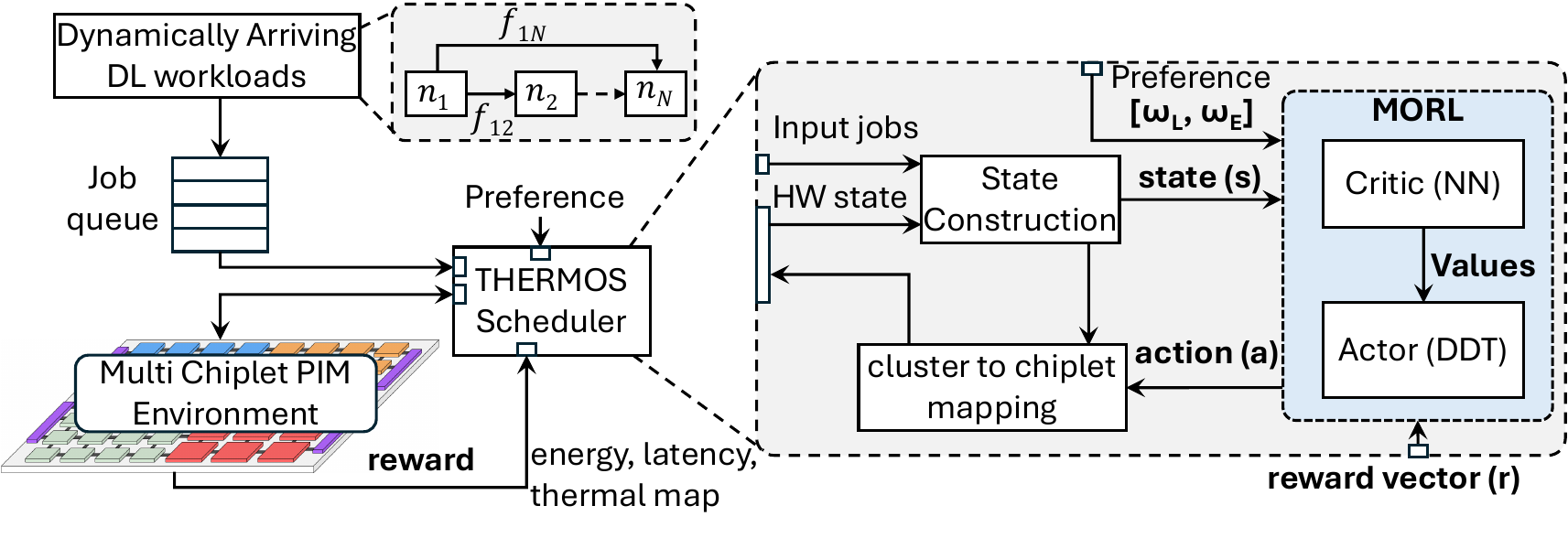}
        \vspace{-15pt}
        \caption{}
        \label{fig:rl_integration}
    \end{subfigure}
    \begin{subfigure}[b]{0.25\textwidth}
        \includegraphics[width=1\linewidth]{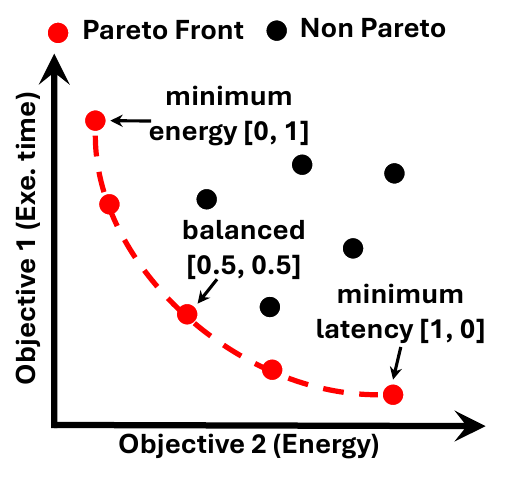}
        \vspace{-15pt}
        \caption{}
        \label{fig:mo_cartoon}
    \end{subfigure}
    \caption{High-level overview of the THERMOS framework. (a) Integration of THERMOS framework with a multi-chiplet PIM environment.  
    (b) Example Pareto front between energy and execution time.} 
    \vspace{-15pt}
    \label{fig:rl_integration}
\end{figure}

\subsection{Overview of the Proposed THERMOS Framework} \label{ssec:overview}
Figure~\ref{fig:rl_integration} provides an overview of the THERMOS framework before we delve into the details in Section~\ref{sec:approach}. 
DL workloads are dispatched by the host processor to the heterogeneous multi-chiplet PIM architecture for inference. 
A newly launched DL model waits in a job queue before THERMOS schedules its neural layers onto the available chiplets.
Optimally scheduling these tasks onto a multi-chiplet PIM system is a challenging sequential decision-making problem. The scheduler must consider dynamically varying constraints such as chiplet availability and temperature. 
Furthermore, the scheduler must meet multiple conflicting objectives, such as latency, energy, power, and thermal constraints, making multi-objective optimization essential.
To achieve this goal, THERMOS captures the characteristics of the incoming tasks (e.g., computation, memory, and communication requirements) and underlying hardware (e.g., available chiplets, memory, and communication resources) and constructs a state space using these features. 
Next, it uses a hierarchical approach to solve this dynamic scheduling problem. 

\bh{Level 1:} Our multi-objective RL (MORL) policy selects \textit{one of the chiplet clusters} to run each neural layer of the current DL workload. Dynamic conditions, such as the temperature of the system and the memory availability of chiplets, are passed as inputs to the RL policy. Static conditions, such as the energy efficiency and performance of each chiplet, are learned by the RL policy during training. The RL policy uses this learned information to make informed mapping decisions. RL is used since it can adapt to changing workloads and system constraints in real-time. However, traditional RL approaches handle multi-objective optimization
by constructing a single objective as a weighted sum of the target objectives or training a separate policy for each objective~\cite{hayes2022practical,van2013scalarized,xu2020prediction,mossalam2016multi}.
Moreover, they suffer from elevated training time and memory requirements, as each policy must be trained and stored individually~\cite{basaklar2022pd,yang2019morl}. 
Therefore, THERMOS trains \textit{a single MORL policy} for multi-objective optimization. 
It uses a preference vector $ \mathbf{\omega} \in \Omega $, where the preferences add up to one, $ \sum_{\mathbf{\omega} \in \Omega } \omega_i = 1 $.
During training, each episode runs all preference vectors in parallel within separate environments, collecting \textit{reward vectors} (not a scalar weighted sum).
These reward vectors are then used to train a single preference-driven policy.
The trained policy takes \textit{an arbitrary preference vector} as input and 
achieves Pareto-optimal objectives as shown in Figure~\ref{fig:mo_cartoon}.

\bh{Level 2:} Once a chiplet type is selected, THERMOS maps the layer across physical chiplets of that type. This stage also includes a tiling mechanism to handle large neural layers that do not fit into a single chiplet's local memory. 
A proximity-driven algorithm first fills one chiplet as much as possible, minimizing inter-chiplet communication. 
If the size of the layer exceeds that chiplet's memory capacity, THERMOS selects additional chiplets of the same type and continues the mapping until the layer is fully assigned. 

In summary, \textit{the proposed framework allows the scheduling policy to be configured at runtime for different target objectives and preferences}. 
We focus on two primary objectives, execution time and energy consumption, while ensuring that the operating temperature of each chiplet remains below its maximum temperature constraint.

\section{Multi-Objective DL Workload Scheduling on Heterogeneous Multi-Chiplet Architectures}\label{sec:approach}
This section begins with a formal definition of the problem and introduces the state and action spaces used in this work. 
Then, it presents the proposed multi-objective DL neural layer to the chiplet cluster scheduling technique and the second-level proximity-driven scheduling algorithm within each cluster.

\vspace{-5pt}
\subsection{Problem Formulation}\label{ssec:problem}
Suppose a new DL model is launched to run a series of inference tasks (e.g., ResNet50 is launched to process a set of images $I$).
Our goal is to find a scheduling policy $\Psi(\mathbf{\omega}): G_{\text{DCG}}(\mathcal{N}, \mathcal{F}) \rightarrow G_{\text{ACG}}(\mathcal{C}, \mathcal{L})$ that schedules the neural layers $\mathcal{N}$ to the chiplets $\mathcal{C}$ to co-optimize the objectives (latency $\mathbf{L}$ and energy $\mathbf{E}$ per workload) as a function of the preference vector $\mathbf{\omega}$. Formally, the optimization problem is stated as:
\begin{equation}\label{eq:objective}
\begin{aligned}
& \min_{\Psi(\boldsymbol{\omega})} \quad 
  \omega_L \mathbf{L} + \omega_E \mathbf{E} 
&& \text{(Co-optimize latency and energy for a given preference vector)}\\[0.2em]
& \text{s.t.} ~~ 
  \forall\, c_i \in \Psi(G_{\text{DCG}}),\; T_i(t) < T^{\max}_i 
&& \text{(All chiplets satisfy the max temperature constraint)}\\[0.2em]
& \omega_L + \omega_E = 1 
&& \text{(Preferences \textit{given at runtime} add up to one)}\\[0.2em]
& \sum_{n_i \in \mathcal{N}} w_i \; \le \;
  \sum_{c_j \in \mathcal{C}} M_j(t) 
&& \text{(DL model weights fit within available chiplet memory)}
\end{aligned}
\end{equation}
%

\bh{Thermal Implications for Dynamic Workload Scheduling:}
The accuracy of DL workloads degrades with temperature in NVM-based PIM macros. PIM macros store weights as conductance values, making them more susceptible to thermal variations~\cite{liu2019hr,meng2021temperature}.
THERMOS adopts a threshold-based throttling approach to limit the accuracy degradation.  
If a chiplet's temperature $T_i(t)$ exceeds $T^{max}_i$, it is throttled, pausing ongoing DL tasks and preventing new task assignments. 
Traditional voltage and frequency scaling is not feasible in PIM since it would impact the accuracy of the system due to analog operation. Therefore, we throttle the chiplets to reduce the power dissipation while avoiding accuracy degradation.
During this time, chiplets consume only leakage energy to maintain the weights stored in crossbar arrays. 
Execution resumes once $T_i(t)$ falls below $T^{max}_i$, ensuring that temperature-induced accuracy degradation is prevented at the expense of higher execution times. 

ReRAM-based chiplets (standard and accumulator) are assigned a threshold temperature of 330K, aligning with previous research on temperature sensitivity~\cite{liu2019hr},
while SRAM-based chiplets (shared ADC and ADC-less) are given a higher threshold of 358K (corresponding to 85$^\circ$C):
\begin{equation}
T^{max}_i =
\begin{cases}
330\,\text{K}, & \text{if } v_i \in \{standard, accumulator\} \\
358\,\text{K}, & \text{if } v_i \in \{shared ADC, ADC-less\}
\end{cases}    
\end{equation}

\subsection{State and Action Space for RL}\label{ssec:state-space}
THERMOS employs a multi-objective RL policy for scheduling neural layers onto PIM clusters.
The scheduling problem is represented as a Markov decision process (MDP) defined by the tuple 
($\mathcal{S}$, $\mathcal{A}$, $P$, $R$, $ \gamma $), 
where $ \mathcal{S} $ and $ \mathcal{A} $ represent the state space and 
action space, respectively. 
$ P(\mathbf{s}'|\mathbf{s},\mathbf{a}) $ is the state transition probability, which is a function of the environment's dynamics, while
$R_t = \sum_{k=0}^{\infty} \gamma^k r(\mathbf{s}_{t+k}, \mathbf{a}_{t+k}) $ is the cumulative reward function and $ \gamma $ is the discount factor.
To solve the MDP efficiently, we employ an RL policy, in which an agent learns an optimal scheduling policy through interaction with the environment. The RL agent observes the system state and schedules neural layers to clusters sequentially, layer-by-layer.

\subsubsection{State Construction:}
The RL agent's state vector $\mathbf{s}$ is constructed using three types of features: layer, DL workload, and PIM cluster features. (Notation follows Table~\ref{tab:notations}.)
To ensure stable learning, all features are normalized before being used as inputs. 
Below, we provide a detailed description of each component.

\bh{Layer Features:}
Since layers are mapped one at a time, the features of individual layers play a crucial role in capturing their computational and memory requirements. For a given layer $n_i$, we consider the following features:
\begin{itemize} 
    \item $w_i$: The memory requirement of the layer.
    \item $o_i$: The number of MAC operations required to execute the layer.
    \item $\sum_{k=0}^{N} f_{ki}$: The total input activations to layer $n_i$, impacting communication time and energy consumption.
\end{itemize}

\bh{DL workload features:}
These features help the RL agent anticipate future resource requirements and optimize scheduling decisions.
By considering workload characteristics, the agent ensures that initial neural layers are scheduled efficiently so that enough memory and resources are available for later layers.
For scheduling a neural layer $n_i$, the DL features include:  
\begin{itemize} 
\item $N - i$: The number of neural layers yet to be scheduled.
\item $\sum_{j=i}^{N} w_j$: The total memory required for the remaining layers.
\item $\sum_{j=i}^{N} o_j$: The total computational requirement of the remaining layers.
\item $\sum_{k=0}^{N} \sum_{j=i}^{N} f_{kj}$: The total activation volume for the remaining layers.
\item $I$: The total input frames to be processed after scheduling the entire DCG. 
\end{itemize}

\bh{PIM clusters features}:
Hardware features provide the RL agent with information about available resources, enabling informed scheduling. 
Since the RL policy selects a cluster rather than individual chiplets, chiplet-level features are aggregated at the cluster level.  
The key system features considered for scheduling include:
\begin{itemize} 
\item  $\sum\limits_{i \mid v_i = v} M_i(t)$: The total available memory across all chiplets of PIM type $v$ at time $t$.
\item  $\max\limits_{i\mid v_i=v} T_i(t)$: Each cluster's maximum temperature allows the agent to account for thermal constraints.
\item $\psi_{i-1}$: The location where the previous layer $n_{i-1}$ was scheduled. This helps the agent to reason about communication costs for activations from the previous layer. 
\end{itemize}

\begin{table}[t]
\centering
\small
\vspace{-10pt}
\caption{Summary of \colorbox{cyan!10}{workload}, \colorbox{yellow!10}{architecture}, and \colorbox{lightgray!20}{MORL} notations used in this work.}
\vspace{-10pt}
\label{tab:notations}

\begin{minipage}{0.48\linewidth}
\centering
\renewcommand{\arraystretch}{0.85}
\begin{tabular}{@{}cl@{}}
    \toprule
    \textbf{Notation} & \textbf{Definition} \\ \midrule
    \rowcolor{cyan!10} $G_\text{DCG}$ & DL workload characterization graph \\
    \rowcolor{cyan!10} $\mathcal{N}$ & Set of neural layers in $G_\text{DCG}$ \\
    \rowcolor{cyan!10} $\mathbf{N}$ & Total neural layers in a workload $\mathbf{N} = |\mathcal{N}|$ \\
    \rowcolor{cyan!10} $\mathcal{F}$ & Set of activation volume between layers \\
    \rowcolor{cyan!10} $n_i$  & Neural layer $i$ \\
    \rowcolor{cyan!10} $w_i$, $o_i$ & Total weights and MAC ops. of $n_i$  \\
    \rowcolor{cyan!10} $f_{ij}$ & Activation volume from $n_i$ to $n_j$ \\
    \rowcolor{cyan!10} $I$ & Number of inputs in a DL workload\\ \midrule

    \rowcolor{yellow!10} $G_\text{ACG}$ & Architecture characterization graph \\
    \rowcolor{yellow!10} $\mathcal{C}$, $\mathcal{L}$ & Set of chiplets and links in ACG \\
    \rowcolor{yellow!10} $\mathcal{V}$ & Set of PIM clusters\\
    \rowcolor{yellow!10} $c_i$, $v_i$ & Chiplet $i$, its PIM type \\
    \rowcolor{yellow!10} $l_{ij}$ & Communication link between $c_i$ and $c_j$  \\ 
    \bottomrule
\end{tabular}
\end{minipage}%
\hfill
\renewcommand{\arraystretch}{0.85}
\begin{minipage}{0.48\linewidth}
    \centering
    \begin{tabular}{@{}cl@{}}
    \toprule
    \textbf{Notation} & \textbf{Definition} \\ \midrule
    \rowcolor{yellow!10} $M_i^\text{cap}$ & Total PIM memory in $c_i$ \\
    \rowcolor{yellow!10} $T_i^\text{max}$ & Maximum allowed temperature in $c_i$ \\
    \rowcolor{yellow!10} $M_i(t)$ &  Memory availability of $c_i$ at time $t$ \\
    \rowcolor{yellow!10} $T_i(t)$ &  Temperature of $c_i$ at time $t$ \\ \midrule
    
    \rowcolor{lightgray!20} $\Psi$ & Scheduling policy for entire DCG\\
    \rowcolor{lightgray!20} $\psi_i$ & Scheduling policy for $n_i$\\
    \rowcolor{lightgray!20} $\omega$ & A preference vector\\
    \rowcolor{lightgray!20} $\pi_{\theta}$ & Actor policy with $\theta$ parameters\\
    \rowcolor{lightgray!20} $\mathbf{V}_{\phi}$ & Critic policy with $\phi$ parameters\\
    \rowcolor{lightgray!20} $\mathbf{s}$ & Input state for MORL \\
    \rowcolor{lightgray!20} $\mathbf{r}$ & Reward vector\\
    \rowcolor{lightgray!20} $\mathbf{a}$ & RL action\\
    \rowcolor{lightgray!20} $\gamma$ & Discount factor\\ 
    
    \bottomrule
\end{tabular}
\end{minipage}

\vspace{-15pt}
\end{table}

\vspace{-5pt}
\subsubsection{Action Space}\label{sssec:action_space}
THERMOS schedules each neural layer sequentially using a discrete action space. 
If a neural layer is too large to fit into a single cluster, the RL agent makes a subsequent decision for the remaining portion of the layer.
In the system considered in this work, multiple chiplet types exist, meaning the RL agent must select one of these types. 
If more or fewer chiplet types existed, the scheduling approach would be identical, and only the training process, which would now include more or fewer chiplet types, would change. 
If the system contained only a single chiplet type, the RL agent's decisions would be trivial, as it would always assign tasks to the only available chiplet type.
In this formulation, the action space, $\mathcal{A}$, is a vector with a length equal to the number of clusters. 
Each element represents the probability of selecting the corresponding cluster. 
The RL agent then chooses the cluster using the \textit{argmax} operation, effectively framing the problem as a categorical classification task. 
This step-by-step mapping leverages the fact that, in most cases, a cluster’s memory exceeds a neural layer’s requirements, thereby reducing unnecessary inter-chiplet communication. 
A continuous action space could also be used to assign layers to different chiplets in a single step. 
However, a continuous action space would make the output space exploration much harder, i.e., computationally expensive and time-consuming, due to the exploding number of possible actions~\cite{sutton2018reinforcement,zhu2021overview}.

\bh{Masking Invalid Actions:}
Occasionally, especially during the early stages of training, the RL agent may select an action where the selected cluster has no available memory. 
We apply invalid action masking instead of penalizing the agent with a large negative reward. This involves assigning a large negative value (-10$^{7}$) to an invalid action before applying the softmax operation to the agent's output, preventing the agent from selecting invalid actions~\cite{huang2022closer}.
The primary advantage of invalid action masking is that it directly restricts the agent from selecting infeasible clusters, ensuring more efficient exploration and stable learning. 
This approach simplifies the reward structure, avoids the complexity of designing and tuning negative rewards, and helps the agent focus on valid clusters.
Additionally, it allows for faster learning by preventing wasted exploration on invalid actions.

\vspace{-5pt}
\subsection{MORL Learning Dynamics}\label{ssec:rl_learning}


We use the actor-critic model, where the actor selects an action $\mathbf{a}$ based on the policy $\pi(\mathbf{a}|\mathbf{s}, \omega)$, 
and the critic evaluates the expected outcome using the value function. 
For a multi-objective problem, the value function is vectorized into $\mathbf{V}(\mathbf{s}, \mathbf{\omega})$, which allows learning each objective individually and generates a Pareto front using the preference vector $\omega$. 
A common approach is to approximate the value and policy functions using neural networks. 
We follow the same approach for the value function, $\mathbf{V}_{\phi}(\mathbf{s}, \omega)$, approximating it with a neural network consisting of three fully connected layers, since it is only used during training. 
The actor policy, however, is implemented using a decision tree, as described next.

\subsubsection{Differential decision tree (DDT) for the RL agent:}
\begin{figure}[b]
\centering
    \begin{subfigure}[b]{0.3\textwidth}
        \includegraphics[width=\linewidth]{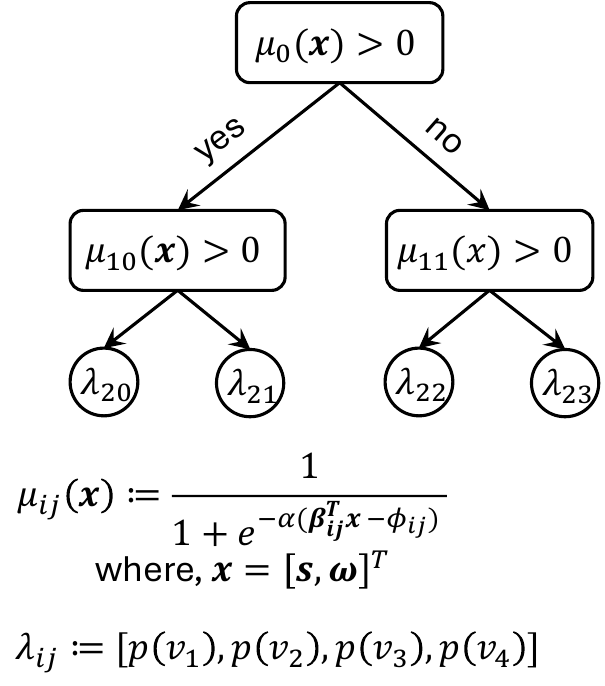}
        \caption{}
        \label{fig:ddt}
    \end{subfigure} 
    \hfill
    \begin{subfigure}[b]{0.6\textwidth}
        \includegraphics[width=\linewidth]{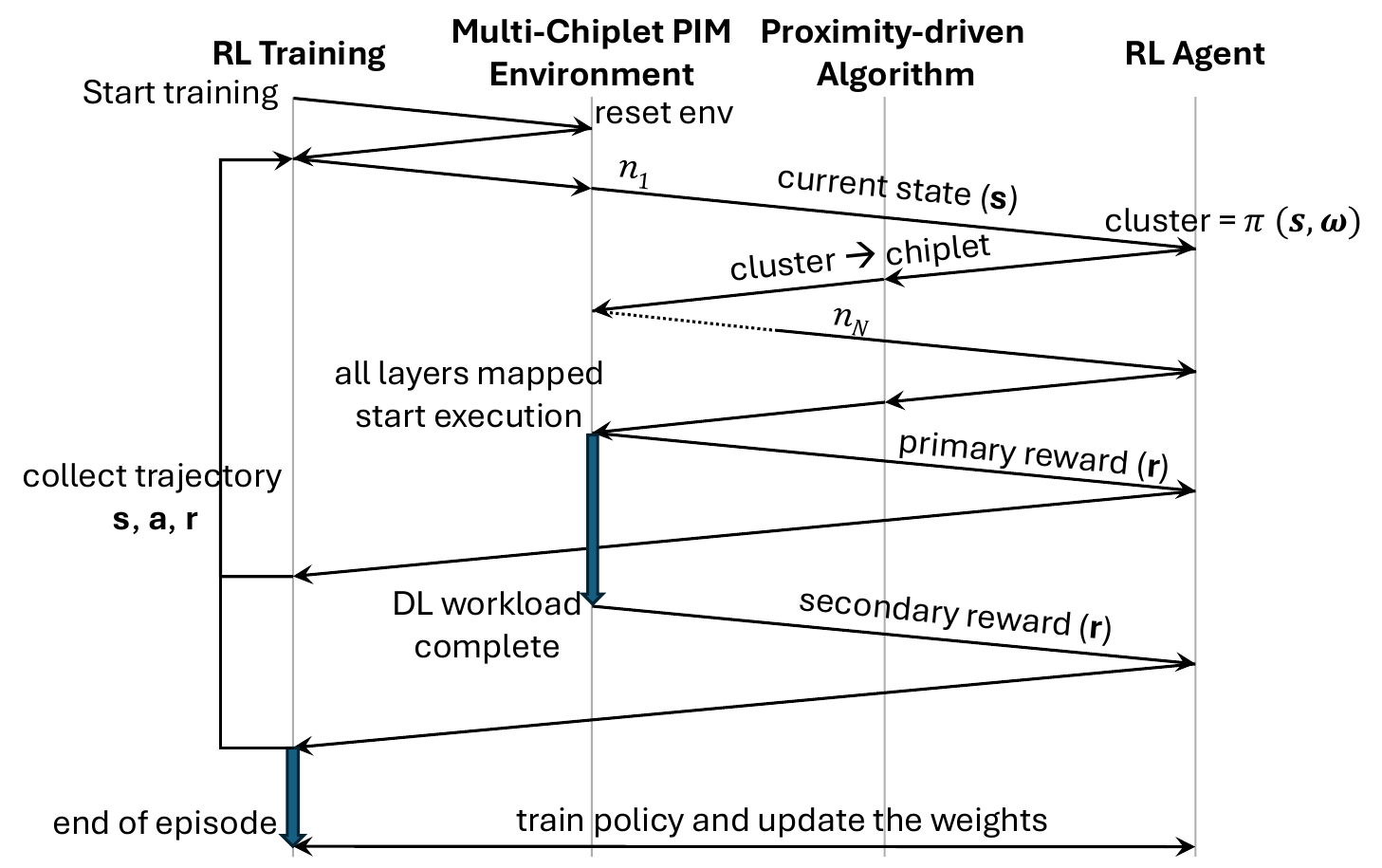}
        \caption{}
        \label{fig:rl_training}
    \end{subfigure}
    \vspace{-10pt}
    \caption{(a) Architecture of DDT of depth 2 ($\alpha, \mathbf{\beta}$ and $\phi$ are learnable parameters and $p(v)$ is the probability of cluster $v$). (b) Interaction between the RL agent and multi-chiplet PIM environment during training.} 
    \vspace{-15pt}
    \label{fig:ddt_training}
\end{figure}

THERMOS implements the RL agent $\pi_{\theta}(\mathbf{a}|\mathbf{s}, \omega)$ using a novel DDT policy since it is explainable and more efficient than a neural network, as demonstrated in Section~\ref{ssec:overhead}.
Unlike traditional decision trees that make binary decisions at each node level, the nodes in the DDT evaluate a linear combination of all state vector ($\mathbf{s}$) components, combined with a preference vector ($\omega$), followed by a sigmoid activation function, as illustrated in Figure~\ref{fig:ddt}. 
This structure enables the modeling of complex, non-linear decision boundaries.
The leaf nodes of the DDT output a vector of size equal to the action space. Applying the softmax function to this vector yields a probability distribution over possible actions, facilitating stochastic action selection, as explained in Section~\ref{sssec:action_space}.
Since this decision tree representation is differentiable, it can be efficiently trained using gradient descent and easily integrated with RL training algorithms~\cite{rodriguez2020optimization}.

\subsubsection{MORL Training Flow:}\label{ssec:morl_training}

We adapt the Proximal Policy Optimization (PPO) algorithm to train MORL~\cite{PPO}. PPO utilizes the advantage function, which can be vectorized given a reward vector ($\mathbf{r}$) and a value function vector $\mathbf{V}_{\phi}(\mathbf{s}, \omega)$ as:
\begin{equation}
\mathbf{A}(\mathbf{s}, \mathbf{a}, \omega) = \mathbf{r} + \gamma \mathbf{V}_{\phi}(\mathbf{s}_{t+1}, \omega) - \mathbf{V}_{\phi}(\mathbf{s}_t, \omega)
\end{equation}
The policy loss for the DDT can be computed as:
\begin{equation}\label{eq:policy_loss}
\begin{aligned}
&L(\theta) = \underset{\mathbf{s}, 
\mathbf{a} \sim \pi_\theta}{\mathbb{E}} \left[ \min \left( r_t(\theta) \omega^T \mathbf{A}(\mathbf{s}, \mathbf{a}, \omega), \, \text{clip}(r_t(\theta), 1 - \epsilon, 1 + \epsilon) \omega^T \mathbf{A}(\mathbf{s}, \mathbf{a}, \omega) \right) \right] 
\text{, where } 
r_t(\theta) = \frac{\pi_{\theta}(\mathbf{a}_t | \mathbf{s}_t, \omega)}{\pi_{\theta_{\text{old}}}(\mathbf{a}_t | \mathbf{s}_t, \omega)}  
\end{aligned}
\end{equation}
In this equation, $\omega^T \mathbf{A}(\mathbf{s}, \mathbf{a}, \omega)$ scalarizes the advantage vector, and $\epsilon$ is the clipping factor.
The loss function for the value network is computed using the mean squared error, which is given by:
\begin{equation}\label{eq:value_loss}
L(\phi) = \underset{\mathbf{s}, 
\mathbf{a} \sim \pi_\theta}{\mathbb{E}} \left[ \| \mathbf{V}_{\phi}(\mathbf{s}_t, \omega) - (\mathbf{r} + \gamma \mathbf{V}_{\phi}(\mathbf{s}_{t+1}, \omega)) \|^2 \right]
\end{equation}
where $\mathbf{r} + \gamma \mathbf{V}_{\phi}(\mathbf{s}_{t+1}, \omega)$ represents the temporal difference (TD) target.
The goal is to train the parameters of the value network $\mathbf{V}_{\phi}(\mathbf{s}, \omega)$ so that it predicts the TD target as accurately as possible for the given preference vector $\omega$.
In this paper, we learn a single DDT with three different preferences: minimizing execution time [1, 0], minimizing energy [0, 1], and balancing both objectives [0.5, 0.5]. 

The RL agent is trained using episodic batches. Each episode begins by resetting the multi-chiplet PIM environment with a randomly selected target throughput, allowing the agent to learn across varying workload demands.
The RL agent selects an action based on the given state and preference vector for each layer in the current DL model, as illustrated in Figure~\ref{fig:rl_training}. 
This action determines a cluster, which is then mapped to a chiplet using a proximity-driven algorithm as described in Section~\ref{ssec:proximity}. 
This process continues until all layers of the current DL model are mapped.  
Then, a reward vector is computed based on the objective functions, and the agent proceeds to the next DL model in the queue. 
The episode continues for a predefined number of DL workloads and ends once all are executed. 
To efficiently train multiple objectives, we use multi-threading to run all three preferences (minimizing execution time, minimizing energy, and balancing both) in parallel. 
Each process interacts with its own copy of the environment and collects trajectories in the form of (state, action, reward) tuples for its respective preference. 
At the end of the episode, these trajectories are used to update the actor-critic models using the loss functions described in Equations~\ref{eq:policy_loss} and ~\ref{eq:value_loss}.  

\subsubsection{Reward Vector Calculation}\label{ssec:reward}
The RL agent's reward is represented as a vector consisting of the negative values of the total execution time and total energy consumed by the DL workload.
Since the RL algorithm aims to maximize the reward, this approach ensures that the reward aligns with the optimization objectives. 
Negating the actual objectives (execution time and energy) incentivizes the agent to minimize the actual objectives. 
To ensure that both objectives are comparable, we normalize and balance the reward values, allowing the agent to learn a Pareto-optimal policy based on the preference vector.

Parallel execution of multiple DL model instances and dynamic effects like thermal throttling complicate reward calculation. THERMOS addresses these challenges with novel strategies detailed below.

\begin{figure}[b]
\centering
    \centering
    \includegraphics[width=0.85\linewidth]{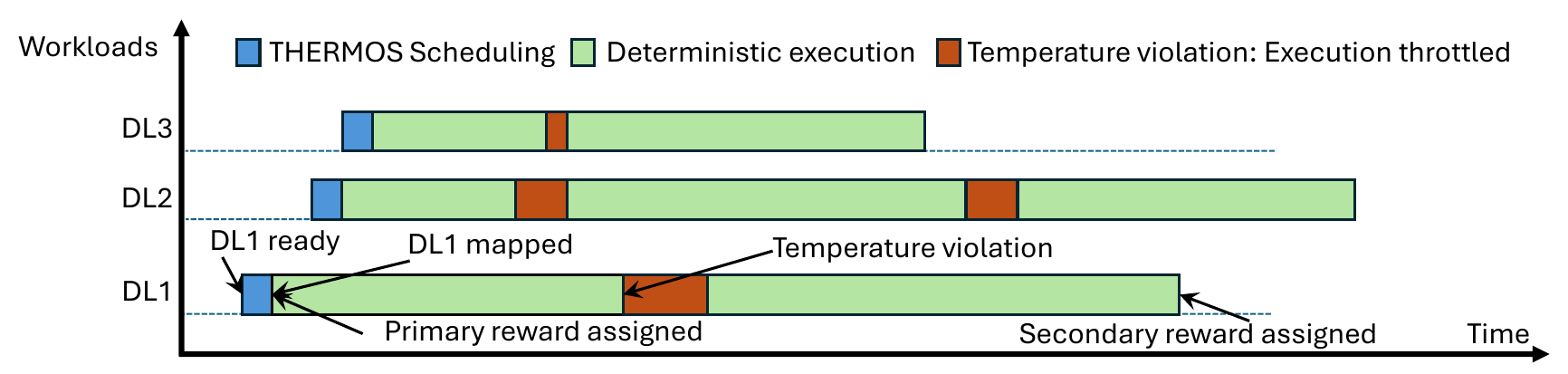}
    \vspace{-10pt}
    \caption{With no temperature violations, deterministic runtime and energy serve as primary rewards, linking rewards to recent actions. When violations occur, extra stall time and energy overhead are assigned as secondary rewards post-execution. Meanwhile, the RL agent continues scheduling other DL workloads, which may be completed in an asynchronous order.} 
    \vspace{-10pt}
    \label{fig:reward_splitting}
\end{figure}

\bh{Tracking delayed rewards:} The RL agent makes sequential decisions for each neural layer. Inference begins only after the entire DL model is mapped. 
Therefore, rewards can only be assigned after the complete workload finishes, resulting in a \textit{delayed reward}, 
while the intermediate actions receive zero rewards.  
However, during the inference of the scheduled workload, new DL workloads may arrive, as the multi-chiplet PIM system can process multiple workloads simultaneously. 
This overlapping of executions leads to asynchronous reward assignment, where rewards are not directly tied to the actions that caused them.
Figure~\ref{fig:reward_splitting} illustrates this issue, showing three DL workloads arriving in the sequence $1 \!\to\! 2 \!\to\! 3$, but completing in the order of $3\!\to\! 1 \!\to\! 2$, leading to delays and misalignment in reward assignments. 
This asynchronous and delayed reward assignment complicates the RL agent’s learning process, as it must attribute outcomes to earlier actions. 
To address this challenge, we propose splitting the reward into two components to improve the agent's ability to learn and optimize the objectives.

\bh{Reward Splitting:}
Suppose the RL agent makes a successful decision under the conditions at time $t_0$, but a thermal throttling caused by later decisions increases the execution time and energy consumption. 
A lumped delayed reward may incorrectly penalize a successful decision. 
To address this issue, we split the reward into two components: primary and secondary.
The primary reward reflects the deterministic execution and is assigned once the entire DL workload is scheduled, assuming ideal conditions. 
This immediate reward helps the RL agent understand the correlation between actions and outcomes.
After the execution is completed, the secondary reward is provided, accounting for the non-deterministic effects, temperature-induced throttling, and the extra energy consumed due to leakage.
Figure~\ref{fig:reward_splitting} illustrates this process, where the primary reward is assigned as soon as DL workload 1 is mapped, and the secondary reward, representing the cumulative effects of temperature violations, is assigned at the end of the execution. 
Although the secondary reward is still asynchronous to the actions, RL algorithms are trained to maximize cumulative rewards. 
This enables the RL agent to learn to minimize both the primary and secondary rewards, effectively reducing execution time and energy consumption over time.

\vspace{-5pt}
\subsection{Proximity-driven Algorithm}\label{ssec:proximity}
After the MORL policy selects a cluster for a neural layer,
a proximity-driven algorithm assigns it to a set of chiplets within the selected cluster. This algorithm minimizes the communication cost since all chiplets in a PIM cluster have identical compute capabilities.  
Since each neural layer in a DCG primarily transmits activations to the subsequent layer (making \( G_{\text{DCG}} \) largely linear), our strategy focuses on reducing inter-chiplet communication overhead. 
To this end, we first identify the chiplets used in the previous layer ($\psi_{i-1}$) and compute the weighted distance from these chiplets to each chiplet in the destination cluster (selected by the MORL policy).
We then sort the chiplets that have available memory (i.e., $M_i(t) > 0$) based on this weighted distance, prioritizing those closest to the previous layer’s chiplets.
Weights are allocated iteratively, filling each chiplet to its capacity before moving to the next. 
This approach minimizes inter-layer communication costs while optimizing memory utilization within the cluster. 
Algorithm~\ref{alg:mapping} details the full THERMOS framework for scheduling deep learning workloads.

\begin{algorithm}[h]\
\caption{THERMOS algorithm: Scheduling DL Workload onto Heterogeneous Multi-Chiplet PIM Architecture}
\label{alg:mapping}
\begin{algorithmic}[1]
\STATE \textbf{Input:} DL workload $G_{\text{DCG}}=(\mathcal{N},\mathcal{F})$, total inputs $I$, architecture graph $G_{\text{ACG}}=(\mathcal{C},\mathcal{L})$, and optimization preference $\omega \in \{\text{energy, latency, balanced}\}$.
\STATE \textbf{Output:} Scheduling policy $\Psi$ from $G_{\text{DCG}}$ onto $G_{\text{ACG}}$.

\STATE \textbf{Initialize:} Current memory availability $\sum\limits_{i|v_i=v} M_i(t)$ and maximum temperature $\max\limits_{i|v_i=v} T_i(t)$ per cluster.

\IF{$\sum\limits_{n_i \in \mathcal{N}} w_i < \sum\limits_{i \in \mathcal{C}} M_i(t)$}  \label{st:check}
  \FOR{each neural layer $ n_i \in \mathcal{N} $}
    \STATE $\text{totalRemainingWeights} \gets w_i$
    \WHILE{$\text{totalRemainingWeights} > 0$} \label{st:while}
      \STATE $cluster \gets \text{\textbf{MORL}}\Big(n_i, \psi_{i-1}, G_\text{ACG}, G_{\text{DCG}}, I, \omega\Big)$ \textbf{// Select a pim-cluster}
      \STATE $\psi_i \leftarrow \text{Proximity-driven algorithm} (cluster, n_i, \psi_{i-1})$  \textbf{// Cluster to Chiplet mapping}
      \STATE Update: totalRemainingWeights, $M_i(t)$  $\forall c_j \in \mathcal{C}$
    \ENDWHILE
  \ENDFOR
  \STATE Aggregate the layer scheduling functions: $\Psi \gets \bigcup\limits_{i} \psi_i $
\ENDIF
\end{algorithmic}
\end{algorithm}

\vspace{-4pt}
\section{Experimental Evaluation} \label{sec:experimental_results}

\subsection{Experimental Setup} \label{ssec:experimental_setup}
\noindent\textbf{Target Architecture: }
We perform extensive experimental evaluations on a 2.5D heterogeneous chiplet-based architecture using four PIM chiplet clusters, as summarized in Table~\ref{tab:pim_types}. 
In addition to these compute chiplets, I/O chiplets are placed at the boundary to load the DL model weights and interface with a host system.

The proposed THERMOS framework is designed to work with any heterogeneous chiplet configuration. This includes systems with any number of chiplets, chiplet clusters, and chiplets per cluster. 
To demonstrate the framework on a practical system, we determine the number of chiplets of each PIM type that co-optimizes system performance and energy efficiency using a multi-objective optimization technique aligned with our runtime objectives~\cite{coello2007evolutionary}. 
The resulting architecture has 25 Standard, 28 Shared ADC, 10 Accumulator, and 15 ADC-less PIM chiplets, as summarized in Table~\ref{tab:pim_types}. 
\textit{We emphasize that the THERMOS framework can work with any configuration.}

\begin{table}[h]
\centering
\vspace{-3mm}
\caption{Properties of different PIM chiplets used to create the heterogeneous multi-chiplet system}
\vspace{-3mm}
\label{tab:pim_types}
\resizebox{\columnwidth}{!}{%
\begin{tabular}{@{}c|ccccccc@{}}
\toprule
\textbf{PIM type} & \textbf{Fabrication tech} & \textbf{Crossbar size} & \textbf{bits/cell} & \textbf{ADC precision} & \textbf{Memory per chiplet} & \textbf{Chiplet Area} & \textbf{\#chiplets} \\ \midrule
\textbf{Standard} & ReRAM & 128$\times$128 & 2 & 8 & 9568 Kb & 4 mm$^2$ & 25 \\
\textbf{Shared ADC} & SRAM & 768$\times$768 & 1 & 8 & 9792 Kb & 9 mm$^2$ & 28 \\
\textbf{Accumulator} & ReRAM & 256$\times$256 & 2 & 8 & 19200 Kb & 4 mm$^2$ & 10 \\
\textbf{ADC-less} & SRAM & 128$\times$128 & 1 & NA & 2416 Kb & 4 mm$^2$ & 15 \\ \bottomrule
\end{tabular}%
}
\vspace{-3mm}
\end{table}

\begin{figure}[t]
\centering
    \begin{minipage}{0.5\linewidth}
    \includegraphics[width=1\linewidth]{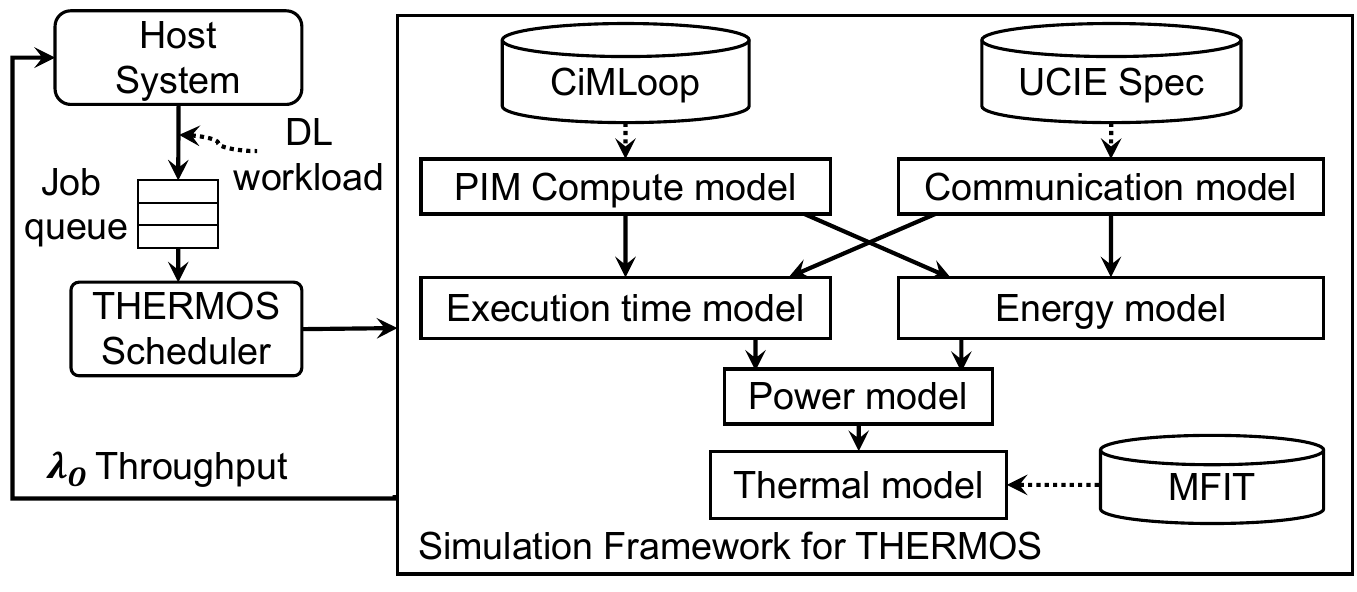}
    \vspace{-6mm}
    \caption{Simulation framework for heterogeneous multi-chiplet PIM systems, modeling DL workloads to evaluate latency, energy, and thermal behavior. Processed outputs are sent to the host with a throughput of $\lambda_O$.} 
    \label{fig:sim}
    \end{minipage}
    \begin{minipage}{0.49\linewidth}
        \centering
    \vspace{-3mm}
    \small
    \captionof{table}{Training and Simulation parameters in THERMOS}
    \label{tab:params}
    \vspace{-3mm}
    \begin{tabular}{@{}l|c@{}}
        \toprule
        Definition & Value \\ \midrule
        Depth of DDT ($\pi_\theta$) & 5 \\
        Layers in critic network ($\mathbf{V}_\phi$) & 3\\ 
        Neurons per layer in critic network & 64\\
        Discount factor ($\gamma$) & 0.95 \\ 
        Learning rate & 5$\times 10^{-4}$ \\
        Job queue size & 20 \\ 
        Warm-up period & 1 minute\\
        Inter chiplet link width & 64 \\
        Communication energy per hop~\cite{sharma2022universal} & 0.5 pJ/b \\
        \bottomrule
    \end{tabular}
    \end{minipage}
\vspace{-10pt}
\end{figure}

\bh{Simulation Methodology: }
The simulation methodology adopted in our evaluations is depicted in Figure~\ref{fig:sim}. The host system dispatches incoming DL workloads into a first-come-first-serve (FIFO) job queue. 
The proposed THERMOS framework dequeues each model before passing it to the RL agent. The RL agent selects the chiplet type to map to before handing off the model to the proximity-driven algorithm, which maps the model to chiplets of that type. 
Models are mapped continuously until the queue is empty or there are insufficient resources to map the next model. 
The host system stalls if the queue is full, while jobs wait until resources are sufficient.
The time from when the host system admits a job into the FIFO queue to when the inference results are returned to the host system is denoted as the end-to-end latency, which also includes the wait time in the job queue. In contrast, execution time refers to the duration from when the job starts execution in the PIM system until it is completed.
To study the steady-state behavior accurately, we use a 1-minute warm-up period and report the average results of ten random simulations, as listed in Table~\ref{tab:params}, among other simulation parameters.

The rest of the simulation framework mimics the hardware operation. 
We simulate the computations within the chiplets using CiMLoop~\cite{andrulis2024cimloop}.
Once THERMOS schedules a neural layer to a chiplet, CiMLoop takes the chiplet and neural layer parameters as inputs. After the simulation, it outputs the corresponding execution time, processing energy consumption, and power dissipation.  
Similarly, we model the inter-chiplet communication due to activations produced by all active chiplets using a custom NoI model and UCIe parameters~\cite{sharma2022universal}. 
Then, we feed the processing and communication power consumption into the open-source MFIT thermal modeling tool to estimate the runtime temperature of each chiplet~\cite{pfromm2024mfit}. We emphasize that this tool uses the floorplan and package geometry, in addition to the power consumption, to model the thermal behavior of the system. 
The details of the simulation framework used in this paper are provided in a GitHub page\footnote{The simulation framework details are available at: \url{https://alishkanani.github.io/THERMOS}. \\ The code repository: \url{https://github.com/AlishKanani/THERMOS}}.

\bh{Training Setup and Convergence Results:} We run three parallel environments, one per preference vector. Each environment collects a trajectory with 10,000 (state, action, reward) tuples, resulting in 30,000 samples per policy update cycle. 
This process proceeds for 25 million environment steps, yielding roughly 8.3 million samples per environment and 833 policy-update cycles in total. 
The optimization follows the modified PPO framework described in Section~\ref{ssec:rl_learning}, a learning rate of $5 \times 10^{-4}$, a clipping threshold $\epsilon = 0.1$, and a discount factor $\lambda = 0.95$. These hyperparameters are held constant across all experiments.
All experiments finish in approximately 5.5 hours for end-to-end training on a single AMD Ryzen Threadripper PRO 7985WX workstation at 3.2GHz clock frequency.

\begin{figure}[h]
\centering
    \centering
    \includegraphics[width=0.7\linewidth]{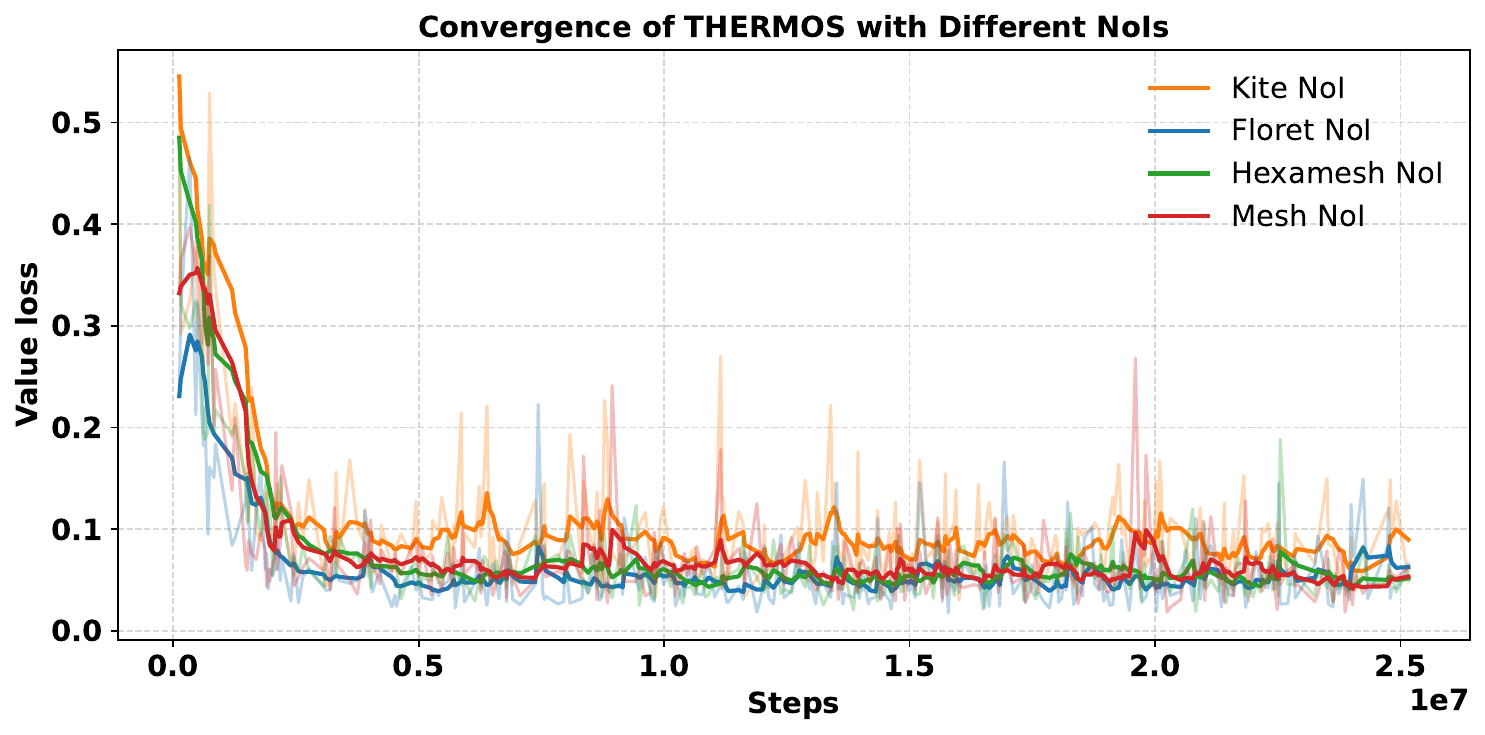}
    \vspace{-12pt}
    \caption{Value-loss versus training steps for four NoI topologies.}
    \vspace{-15pt}
    \label{fig:v_loss}
\end{figure}

Figure~\ref{fig:v_loss}
plots both the raw and exponentially smoothed ($\alpha$ = 0.8) value-loss curves for the Kite, Floret, Hexamesh, and Mesh NoI configurations. All curves plateau below 0.06 after approximately 15 million steps and remain stable for the remainder of training, indicating that the training has converged and the shared actor generalizes without additional tuning.

\subsection{Target DL Workloads and Baseline Scheduling Algorithms}\label{ssec:workload_baseline}
\noindent\textbf{DL workloads: }
Our workloads consist of deep neural network (DNN) models for image processing tasks. These models use weight-stationary layers, requiring fewer rewrites of ReRAM cells, which have limited write endurance. 
Specifically, we employ six commonly used DNN models for PIM research: AlexNet, ResNet18, ResNet50, EfficientNet-B3, MobileNetV3-Large, and Inception-v3. 
These models differ significantly in their number of weights, total MAC operations, and intermediate activations, thus reflecting a broad range of computational characteristics.
To further generalize the workload, we vary the number of images processed by each model, allowing up to 20,000 images per DNN. 
We then create a workload mix of 500 (DNN, \#images) tuples, randomly selecting both the model and the image count to evaluate THERMOS on different multi-chiplet PIM systems connected via Mesh, Kite, Floret, and Hexamesh NoI architectures~\cite{kite,sharma2023florets,hexamesh}.


During the MORL training phase, each episode randomly selects 100 DNNs and a random number of images.
This setup exposes the RL agent to diverse scenarios, improving its ability to generalize across different DNN types and workload sizes.
Suppose we want to train a policy for three preferences: [1,0] (latency-minimization), [0,1] (energy-minimization), and [0.5,0.5] (balanced). 
We launch three parallel simulator instances for each training episode, one for each preference vector. The resulting state–action–reward trajectories are pooled to update a single preference-driven actor–critic policy $\pi(\textbf{a}|\textbf{s}, \omega)$. 
Since all three parallel threads run concurrently, the training time is similar to that of a single-environment setup. 
Our multi-objective RL technique generally samples multiple random preference vectors for training and can generalize to any preference vector at deployment.

\bh{Baseline scheduling algorithms:}
We compare THERMOS against three baseline schedulers from prior work. 
First, Simba is a chiplet-based approach that primarily schedules consecutive layers in close physical proximity, reducing communication overhead~\cite{simba}.
However, it focuses solely on minimizing data transfer. 
Second, using a single PIM type, Big-Little chiplets explore heterogeneity in chiplet size rather than PIM type. Big chiplets have larger crossbars than little chiplets~\cite{krishnan2022biglittle}.
Since early DNN layers require less memory than subsequent layers, they are mapped to smaller chiplets, keeping larger ones available for later layers. Scheduling is based on crossbar utilization, assigning workloads to chiplets with the highest utilization.
We adapt this methodology for four different PIM types.
Finally, RELMAS is an RL-based scheduling framework originally designed for heterogeneous accelerators consisting of Simba and Eyeriss-style tiles~\cite{blanco2024deep}.
We adapt it to our context and compare its performance with THERMOS. 
Unlike THERMOS’s fast and lightweight differentiable decision tree and two-level approach with chiplet cluster, RELMAS uses a neural network policy and uses a flat approach to select individual chiplets.

\vspace{-5pt}
\subsection{Evaluation of THERMOS on Mesh NOI}\label{ssec:mesh}
This section comprehensively evaluates THERMOS on a multi-chiplet PIM architecture interconnected via a mesh NoI. 
We train a single DDT policy with a maximum depth of five to schedule the DNN models onto the PIM clusters. 
The trained policy uses the preference vector provided at runtime to optimize for different objectives adaptively. 
Since the heterogeneous chiplet architecture can execute multiple DNNs concurrently, we stream the mix of DNNs at varying arrival rates into this architecture.

\begin{figure}[t]
\centering
    \centering
    \vspace{-10pt}
    \includegraphics[width=0.9\linewidth]{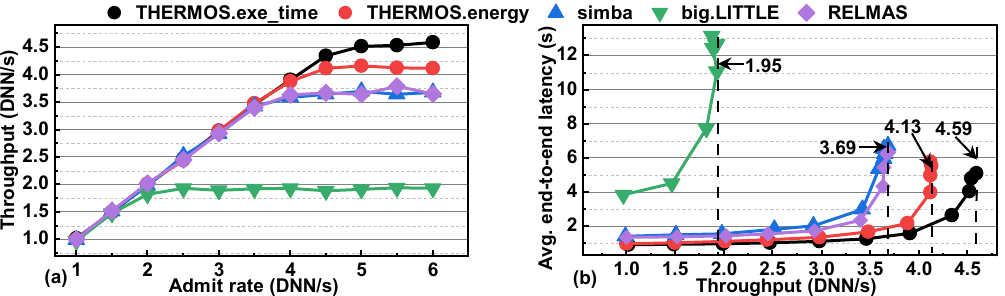}
    \vspace{-10pt}
    \caption{Comparison of THERMOS compared to other baselines with Mesh NoI. (a) Throughput vs. admit rate, (b) end-to-end latency vs. throughput } 
    \label{fig:throughput_results}
\end{figure}

\bh{Maximum achievable throughput:}
As the arrival rate of DNN workloads from the host increases, all scheduling algorithms initially achieve a proportional scaling throughput but begin to saturate at different throughput levels, as shown in Figure~\ref{fig:throughput_results}a. 
Notably, the Big-Little-based scheduler reaches an early saturation point, achieving a maximum throughput of only 1.95 DNNs per second. Simba and RELMAS outperform it, achieving up to 3.69 DNNs per second throughput.
THERMOS significantly outperforms these baselines. 
When the minimum execution time is the only preference ($\mathbf{\omega}=[1,0]$), 
THERMOS achieves the highest throughput of 4.59 DNNs per second.
The same policy reaches 4.43 DNNs per second when the preferences are equal ($\mathbf{\omega}=[0.5,0.5]$), while the minimum energy preference attains 4.13 DNNs per second, which is still higher than the baseline. 
These results demonstrate that THERMOS can effectively balance workload distribution to maximize system utilization.
Figure~\ref{fig:throughput_results}b presents the average end-to-end latency per DNN, which also factors in the waiting time in the job queue for any DNN that cannot be admitted immediately. 
Consequently, all scheduling algorithms exhibit higher latency as throughput increases, since newly arrived DNNs spend more time waiting before they can be scheduled. 
Despite this general trend, THERMOS not only achieves the highest throughput but also maintains a lower end-to-end latency. 
In contrast, the Big-Little scheduler exhibits a sharp increase in latency as throughput grows, highlighting its inefficiency under high load. 
Simba and RELMAS similarly experience rising latency beyond 3 DNNs per second.
Overall, these results emphasize that THERMOS effectively balances throughput and latency, ensuring optimal performance in multi-DNN execution scenarios. 

\begin{figure}[b]
\centering
    \centering
    \includegraphics[width=1\linewidth]{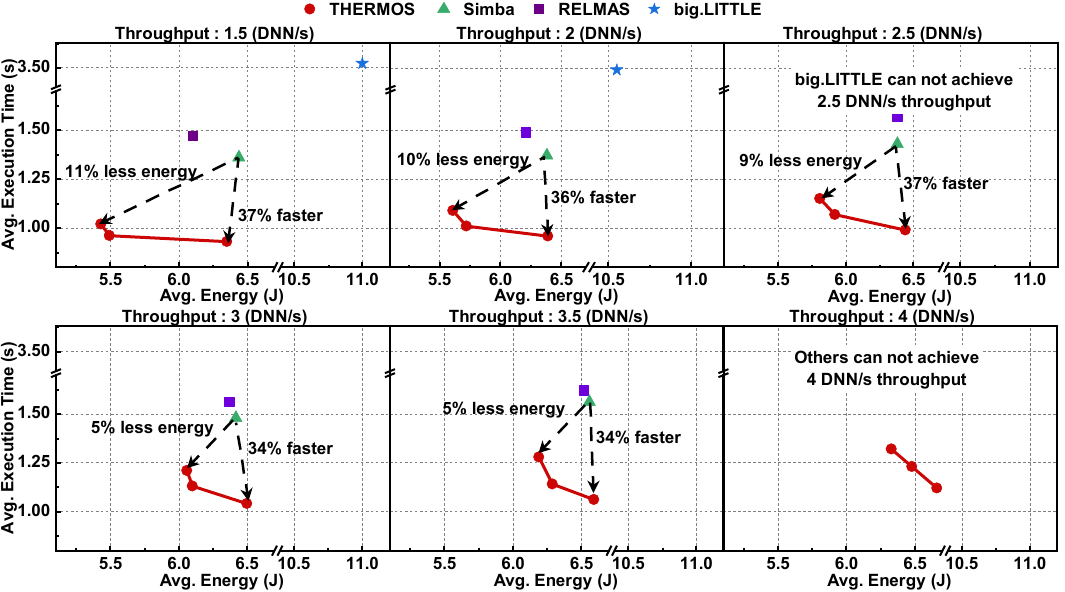}
    \vspace{-15pt}
    \caption{A single Pareto-optimal THERMOS policy and comparison against other scheduling policies under different throughput scenarios for Mesh NoI.} 
    \vspace{-10pt}
    \label{fig:mesh_results}
\end{figure}

\bh{Pareto-optimal execution time/energy consumption results:}
Next, we demonstrate Pareto-optimal results across two key objectives: execution time and energy, obtained with a single THERMOS policy. 
Figure~\ref{fig:mesh_results} shows the average execution time versus average energy for increasing throughput scenarios. 
These plots show the execution time, i.e., the time a DNN is admitted to the multi-chiplet PIM system until it finishes execution, without including waiting time in the job queue.

In the top-left plot of Figure~\ref{fig:mesh_results}, we compare different scheduling policies at a throughput of 1.5 DNN/s. 
Ideally, a policy should be close to the origin, minimizing both execution time and energy. 
The connected dots represent a single THERMOS policy with three preferences, while single points indicate baseline policies. 
Notably, THERMOS achieves Pareto-optimal results and is closest to the origin, demonstrating superior efficiency.
As throughput increases, both execution time and energy rise across all scheduling algorithms. 
At lower throughput, the system is mostly idle, making the most suitable chiplets for a particular layer available. 
When throughput increases, the best chiplets may be occupied, forcing the scheduler to choose suboptimal ones.
Despite these constraints, THERMOS finds Pareto-optimal solutions and outperforms state-of-the-art techniques under diverse workload conditions, as shown by the plots in Figure~\ref{fig:mesh_results}. 
For example, it is 36--37\% faster and 9--10\% more energy efficient than the best-performing baseline (Simba) at 2 and 2.5 DNN/s throughput. 

A key challenge is that selecting the best compute chiplets layer-by-layer in isolation often leads to suboptimal results due to inter-layer communication overhead.
The proposed MORL technique captures these dependencies, allowing it to balance computation and communication costs during training. 
In contrast, Simba, which primarily minimizes communication overhead, and Big-Little, which emphasizes maximizing crossbar utilization, thus focusing on minimizing compute cost, fail to capture this trade-off effectively. 
RELMAS also uses reinforcement learning, but its policy relies on selecting from all chiplets directly, leading to a vast search space that hinders convergence toward optimal solutions.
With its single policy adapting to different runtime preferences, THERMOS outperforms all baselines in execution time and energy. 

When configured to minimize execution time (THERMOS.exe\_time), THERMOS achieves an average speedup of 35\% over Simba, 72\% over Big-Little, and 31\% over RELMAS for similar throughput. 
Big-Little cannot exceed a throughput of 2 DNNs per second, and Simba and RELMAS cannot reach 4 DNNs per second. 
When configured to minimize energy (THERMOS.energy), THERMOS achieves 8\%, 48\%, and 11\% lower energy consumption compared to Simba, Big-Little, and RELMAS, respectively.
To evaluate balanced preference, we use the Energy-Delay Product (EDP). 
THERMOS, with a balanced preference (THERMOS.balanced), reduces EDP by approximately 36\%, 88\%, and 34\% compared to Simba, Big-Little, and RELMAS, respectively. 
Table~\ref{tab:summary} presents the average percentage improvement of THERMOS across all throughput scenarios compared to other scheduling policies.
Since the difference in energy consumption between THERMOS’s minimum execution time and minimum energy preference settings is typically less than half a joule (or under 10\%), more than three distinct preference modes would offer minimal practical benefit. 
Hence, we limit the scope to three preferences in this paper. 
Additionally, unlike other methods that provide only a single operating point, THERMOS offers a \textit{single adaptive policy that can be configured at runtime} to meet varying objectives while remaining Pareto-optimal.
By leveraging its adaptive scheduling policy, THERMOS successfully maximizes system utilization while reducing per-DNN execution time and energy, making it a superior solution for heterogeneous PIM architectures.

\begin{table}[b]
\centering
\vspace{-10pt}
\caption{Average percentage improvement achieved by a single multi-objective THERMOS policy with three preferences, compared to baseline scheduling algorithms.}
\vspace{-10pt}
\label{tab:summary}
\resizebox{\columnwidth}{!}{%
\begin{tabular}{@{}c|ccc|ccc|ccc@{}}
\toprule
\multirow{2}{*}{\textbf{NoI}} & \multicolumn{3}{c|}{\textbf{\begin{tabular}[c]{@{}c@{}}THERMOS.exe\_time \\ \% speedup w.r.t.\end{tabular}}} & \multicolumn{3}{c|}{\textbf{\begin{tabular}[c]{@{}c@{}}THERMOS.energy \\ \% energy reduction w.r.t.\end{tabular}}} & \multicolumn{3}{c}{\textbf{\begin{tabular}[c]{@{}c@{}}THERMOS.balanced \\ \% EDP improvement w.r.t.\end{tabular}}} \\ \cmidrule(lr){2-10} 
 & \textbf{Simba} & \textbf{Big-Little} & \textbf{RELMAS} & \textbf{Simba} & \textbf{Big-Little} & \textbf{RELMAS} & \textbf{Simba} & \textbf{Big-Little} & \textbf{RELMAS} \\ \midrule
\textbf{Mesh} & 35.29\% & 72.00\% & 30.99\% & 8.31\% & 48.33\% & 11.41\% & \textbf{36.41\%} & 87.93\% & 34.40\% \\
\textbf{Floret} & 15.58\% & \textbf{88.83\%}& \textbf{57.93\%} & \textbf{22.17\%} & \textbf{56.52\%} & \textbf{25.67\%} & 26.77\% & \textbf{112.62\%} & \textbf{104.15\%} \\
\textbf{Hexamesh} & 33.37\% & 71.36\% & 34.24\% & 8.83\% & 51.16\% & 13.12\% & 30.83\% & 88.61\% & 34.65\% \\
\textbf{Kite} & \textbf{36.44\%} & 81.20\% & 37.48\% & 2.43\% & 44.77\% & 5.42\% & 23.33\% & 99.46\% & 36.77\% \\ \bottomrule
\end{tabular}%
}
\vspace{-15pt}
\end{table}

\bh{Thermal Constraint Effectiveness:}
We analyze the effectiveness of our thermal management strategy by comparing unconstrained and thermally-constrained scheduling policies. 
Without constraints, the system experiences frequent and sustained thermal violations, leading to prolonged overheating that can compromise reliability and inference accuracy. In contrast, our fine-grained thermal constraint mechanism actively mitigates such violations by pausing execution on any chiplet that exceeds the temperature threshold. This intervention not only cools the overheated chiplet but also slows the thermal rise in adjacent chiplets, preventing widespread overheating. The result is a significant reduction in both the duration and severity of thermal excursions, enabling the system to maintain thermal headroom with minimal disruption to throughput.

\vspace{-5pt}
\subsection{Evaluation of THERMOS for Different NOI Architectures.}
To demonstrate the generalizability of THERMOS, we evaluate its performance on three alternative NoI architectures from the literature (Floret, Hexamesh, and Kite) using the same number of chiplets as in our baseline mesh. 
Our results show that THERMOS adapts effectively to diverse interconnect designs.

\begin{figure}[b]
\centering
    \centering
    \includegraphics[width=1\linewidth]{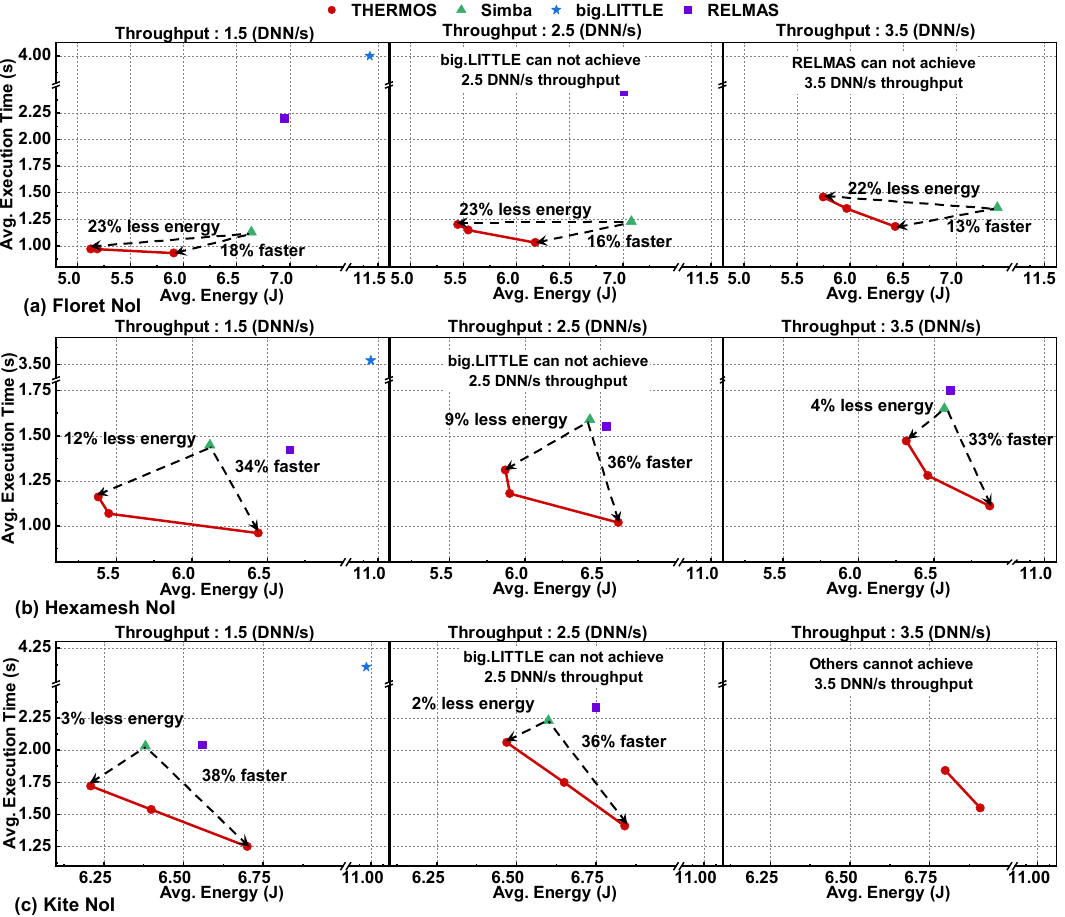}
    \vspace{-15pt}
    \caption{A single Pareto-optimal THERMOS policy and comparison against other scheduling policies under different throughput scenarios for (a) Floret, (b) Hexamesh, and (c) Kite NoI.} 
    \vspace{-15pt}
    \label{fig:other_noi_results}
\end{figure}

Floret is a space-filling curve (SFC) based design that arranges chiplets in a continuous data flow, forming four SFCs, one for each cluster in our implementation~\cite{sharma2023florets}. 
Hexamesh, in contrast, is a 2D staggered topology where each chiplet connects to six neighbors~\cite{hexamesh}. It uses roughly 5\% more interposer area. 
Finally, Kite is a torus-like architecture featuring diagonal skip links to minimize hop counts~\cite{kite}. 
We adopt the kite-small variant that limits skip connections to only nearby diagonals to comply with passive silicon interposer constraints in the UCIe specification (which disallows links longer than 2 mm).
For each NoI architecture, we train a single RL policy with DDT and test it with three preferences: execution time, energy, and a balanced objective. We then compare THERMOS against Simba, Big-Little, and RELMAS, mirroring the setup from our mesh experiments. 
Figure~\ref{fig:other_noi_results} summarizes the results under three throughput scenarios, showing that all schedulers exhibit higher execution time and energy at higher throughput, reflecting increased system utilization and resource contention.

In Floret, THERMOS optimized for execution time achieves 16\% faster execution than Simba, 89\% faster than Big-Little, and 58\% faster than RELMAS on average. 
When configured for minimum energy, THERMOS reduces energy consumption by 22\%, 57\%, and 26\% compared to Simba, Big-Little, and RELMAS, respectively. 
Under the balanced preference, THERMOS achieves 27\%, 113\%, and 104\% lower EDP than these baselines. Notably, Big-Little cannot reach 2.5 DNNs/s, and RELMAS cannot achieve 3.5 DNNs/s throughput.

Similar trends appear in Hexamesh, where THERMOS, with a minimum execution time preference, executes workloads on an average of 33\%, 71\%, and 34\% faster than Simba, Big-Little, and RELMAS, respectively. 
When configured for minimum energy, THERMOS consumes 9\%, 51\%, and 13\% less energy, and under the balanced preference, it reduces EDP by 31\%, 89\%, and 35\%. Again, Big-Little cannot achieve 2.5 DNNs/s in this architecture.

In Kite, THERMOS with minimum execution time preference achieves, on average, 36\%, 81\%, and 37\% faster execution than Simba, Big-Little, and RELMAS. Configured for minimum energy, THERMOS consumes 2\%, 45\%, and 5\% less energy than these baselines and, with a balanced preference, achieves 23\%, 99\%, and 37\% lower EDP. Here, Big-Little cannot reach 2.5 DNNs/s, and RELMAS cannot achieve 3.5 DNNs/s throughput.

Overall, these results confirm that THERMOS’s single adaptive policy maintains its superiority in execution time, energy, and balanced performance objectives, regardless of the underlying NoI architecture. 
By intelligently balancing compute and communication costs, THERMOS attains Pareto-optimal results in all tested topologies, showing its effectiveness and broad applicability for heterogeneous PIM systems.

\vspace{-5pt}
\subsection{Scalability and Overhead Analysis}\label{ssec:overhead}

Scalability and minimal overhead are essential considerations for the practical use of the proposed technique. Therefore, this section thoroughly analyzes the scalability and overhead of implementing the THERMOS scheduler on an NVIDIA Jetson Xavier NX board~\cite{nvidia2020jetson}.

THERMOS is designed as a general framework for thermal-aware scheduling on heterogeneous chiplet-based architectures. 
While evaluations focus on PIM chiplets and explore four representative PIM types, \textit{the scheduling technique itself is agnostic to the specific compute technologies integrated into the package.}
THERMOS is scalable in the following ways. First, it requires only a single policy that works with a preference vector input. Second, the hierarchical scheduler assigns work to clusters of chiplets rather than individual dies. 
Consequently, policy input/output size depends on the number of clusters, not the total chiplets. Our experiments already span around 80 chiplets mapped to 4 clusters, exceeding the scale of any publicly disclosed system. 
Finally, THERMOS represents each neural network layer as a directed graph whose vertices and edges capture MAC intensity and communication volume, respectively. This abstraction enables the trained policy to generalize to different DNN workloads without task-specific retraining.

The MORL policy with the DDT agent and the second-level proximity-driven chiplet allocation algorithm are implemented on an NVIDIA Jetson Xavier NX board~\cite{nvidia2020jetson}. 
Both components are executed one million times to measure the average execution time via software counters. 
We use power data collected via a shunt resistor on a custom PCB for energy measurements, monitoring current and voltage with an NI PXIe-6356 DAQ system~\cite{ni_pxie_6356}.

\begin{figure}[h]
\centering
    \includegraphics[width=0.6\linewidth]{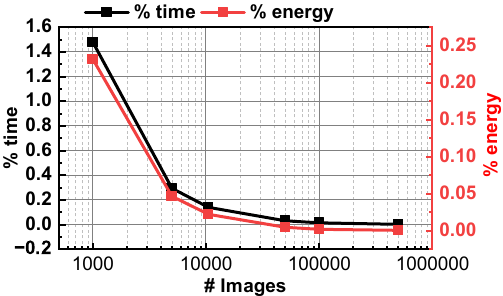}
    \vspace{-12pt}
    \caption{THERMOS overhead analysis with increasing number of images.}
    \vspace{-10pt}
    \label{fig:overhead}
\end{figure}

Table~\ref{tab:overhead} summarizes the execution time and energy consumption overheads. 
Since the MORL policy uses a DDT with five levels, where each node involves only a vector-vector multiplication and a sigmoid function, the time per inference is minimal, averaging 0.6 $\mu s$. 
In contrast, the proximity-driven algorithm requires computing the shortest weighted distance from chiplets of the previous neural layer to chiplets in a particular cluster, consuming 49.3 $\mu s$. 
As a result, the total overhead is 49.9 $\mu s$ per call. 
A similar pattern appears in energy consumption: the DDT requires merely 0.36 $\mu J$ per call, whereas the proximity-driven algorithm consumes 44.37 $\mu J$, yielding a total of 44.73 $\mu J$ per call. 
To quantify how the THERMOS scheduling overhead scales with model execution time, we evaluated the DDT policy and proximity-driven allocation across workloads of 1,000, 5,000, 10,000, 50,000, 100,000, and 500,000 images. 
As shown in Figure~\ref{fig:overhead}, the relative overhead decreases sharply as the number of images increases.
Even for the smallest workload (1,000 images), the total runtime impact remains under 1.5\%, and energy overhead under 0.25\%, demonstrating that the scheduling cost is almost imperceptible in real‐world deployments. 
These results confirm that THERMOS achieves Pareto‐optimal scheduling decisions through a single, low‐overhead RL-based policy, consistently outperforming baseline algorithms in both execution time and energy objectives across a wide range of inference durations.


\begin{table}[t]
\centering
\vspace{-5pt}
\caption{THERMOS overhead analysis on the Nvidia Jetson Xavier NX board~\cite{nvidia2020jetson}}
\vspace{-10pt}
\label{tab:overhead}
\resizebox{\textwidth}{!}{%
\begin{tabular}{@{}c|ccccc@{}}
\toprule
 & \textbf{Time per call} & \textbf{Energy per call} & \textbf{\begin{tabular}[c]{@{}l@{}}\% time per DNN \\ with $\sim$10000 images \end{tabular}}  & \textbf{\begin{tabular}[c]{@{}l@{}}\% Energy per DNN \\ with $\sim$10000 images \end{tabular}} \\ \midrule
\textbf{RL policy} & 0.6 ($\mu s$) & 0.36 ($\mu J$) & 0.0017\% & 0.00018\% \\
\textbf{Proximity-driven algorithm} & 49.3 ($\mu s$) & 44.37 ($\mu J$) & 0.14\% & 0.022\% \\
\textbf{THERMOS (Combined)} & 49.9 ($\mu s$) & 44.73 ($\mu J$)  & \textbf{0.14\%} & \textbf{0.022\%} \\ \bottomrule
\end{tabular}%
}
\vspace{-10pt}
\end{table}

To perform thermal modeling in this work, we employ the open-source MFIT framework~\cite{pfromm2024mfit}, leveraging its variable-granularity node placement to balance accuracy and efficiency. 
Specifically, we map each active chiplet layer to a dense 2$\times$2 node grid, while non-active layers (substrate, lid) use coarser meshes, resulting in a total of 580 thermal nodes. 
We select MFIT’s discrete-state-space (DSS) model for its matrix-vector formulation, which enables very fast simulations. 
By discretizing the model at a 100-ms sampling interval, we achieve temperature monitoring every 100 milliseconds.
On our test platform, each DSS invocation requires approximately 15 $\mu$s, corresponding to 15 $\mu$s per 100 ms of runtime or about 0.015\% of total execution time. 
This fixed, negligible overhead ensures that thermal monitoring adds minimal cost while delivering essential temperature feedback for mapping decisions and ongoing reliability assurance.

\subsection{Limitations and Future Work}\label{ssec:limitations_future_work}

While THERMOS demonstrates significant improvements over state-of-the-art schedulers, future work can address several aspects to improve its practicality and adoption. 
First, the current policy is trained for a representative and practical set of objectives: execution time, energy, and thermal constraints. As additional objectives or finer-grained preference vectors are introduced, the preference-driven PPO algorithm must navigate a higher-dimensional reward landscape, which may impact policy quality and training stability. 
Hence, scaling to larger objective sets (e.g., reliability, cost) represents a possible future research direction.
Second, although the proposed two-level scheduling technique is chiplet technology agnostic, our experimental evaluation focuses on PIM chiplets. Another interesting direction is adapting the power, thermal, and performance models for other compute substrates (e.g., systolic arrays, CPU/GPU tiles, or domain-specific accelerators) and applying THERMOS.
Finally, the current policy training relies on trace-driven simulation. Integrating THERMOS with in-package monitors can enable validations on silicon and evaluations under real workload and process variations.
\section{Conclusion} \label{sec:conclusion}
This work addresses the complex challenge of scheduling deep learning workloads in heterogeneous multi-chiplet PIM architectures, where conflicting objectives such as latency, energy efficiency, and thermal constraints must be balanced at runtime. 
Our proposed solution, THERMOS, employs a hierarchical approach that combines a single reinforcement learning policy that achieves Pareto-optimal results for a preference vector given at runtime with a proximity-driven algorithm. 
This approach enables the scheduler to adapt to varying runtime preferences while dynamically balancing computation and communication costs. 
Experimental results show that THERMOS significantly outperforms existing baselines, delivering substantial speedups and energy reductions under various workload conditions and interconnect architectures. 
Overall, THERMOS provides a robust framework for adaptive, multi-objective scheduling, paving the way for more efficient and resilient AI accelerator designs in next-generation heterogeneous systems.

\bibliographystyle{ACM-Reference-Format}
\bibliography{reference/iarch, reference/spec, reference/related,reference/ogras_paper}




\end{document}